%% file: journalpaper.tex
\documentclass[11pt]{article}

\usepackage[margin=1in]{geometry} 
\usepackage{graphicx} 
\usepackage{amsmath} 
\usepackage{amsfonts} 
\usepackage{amsthm} 
\usepackage{xspace}
\usepackage{subfiles}
\usepackage{blindtext}
\usepackage{mymacros}
\usepackage{amssymb}
\usepackage{bm}
\usepackage{etoolbox}
\usepackage{marginnote}
\usepackage{tikz-cd}
\usepackage{pst-node}
\usepackage{todonotes}

\newtheorem{thm}{Theorem}
\newtheorem{theorem}[thm]{Theorem}

\newtheorem{definition}[thm]{Definition}

\newtheorem{lemma}[thm]{Lemma}

\newtheorem{corollary}[thm]{Corollary}

\begin{document}

\title{Symmetric Arithmetic Circuits\thanks{Research funded by EPSRC
    grant EP/S03238X/1.  A preliminary version of this work was reported in~\cite{DawarW20}}}

\author{Anuj Dawar and Gregory Wilsenach \\
  Department of Computer Science and Technology\\ University of Cambridge.\\
  \texttt{anuj.dawar@cl.cam.ac.uk, gregory.wilsenach@cl.cam.ac.uk}}

\maketitle

\begin{abstract}
\subfile{sections-arxiv/abstract}
\end{abstract}

\section{Introduction}\label{sec:introduction}

\subfile{sections-journal/introduction.tex}

\section{Background}\label{sec:background}

\subfile{sections-journal/background.tex}

\section{Symmetric Circuits}\label{sec:symm-circ}
\subfile{sections-journal/symmetric-arithmetic-circuits.tex}

\section{An Upper-Bound for the Determinant}\label{sec:determinant}

\subfile{sections-journal/determinant}

\section{From Arithmetic To Boolean Circuits}\label{sec:formulas-to-circuits}

\subfile{sections-journal/arithmetic-boolean.tex}

\section{Supports and Counting-Width}\label{sec:supports}
\subfile{sections-journal/supports.tex}

\section{A Lower-Bound for the Permanent}\label{sec:permanent}

\subfile{sections-journal/permanent-lower-bound.tex}

\section{Concluding Discussion}

\subfile{sections-journal/conclusion.tex}

\bibliographystyle{plain} \bibliography{references.bib}
\end{document}

%% file: sections-arxiv/abstract.tex
We introduce symmetric arithmetic circuits, i.e.~arithmetic circuits with a
natural symmetry restriction. In the context of circuits computing polynomials
defined on a matrix of variables, such as the determinant or the permanent, the
restriction amounts to requiring that the shape of the circuit is invariant
under simultaneous row and column permutations of the matrix.  We establish unconditional
exponential lower bounds on the size of any symmetric circuit for computing the
permanent.  In contrast, we show that there
are polynomial-size symmetric circuits for computing the determinant over fields
of characteristic zero.

%% file: sections-journal/introduction.tex
Valiant's conjecture~\cite{Valiant79}, that $\VP\neq\VNP$, is often referred to
as the algebraic counterpart to the conjecture that $\PT\neq\NP$. It has proved
as elusive as the latter. The conjecture is equivalent to the statement that
there is no polynomial-size family of arithmetic circuits for computing the
permanent of a matrix, over any field of characteristic other than 2. Here,
arithmetic circuits are circuits with input gates labelled by variables from
some set $X$ or constants from a fixed field $\ff$, and internal gates labelled
with the operations $+$ and $\times$. The output of such a circuit is some
polynomial in $\ff[X]$, and we think of the circuit as a compact representation
of this polynomial. In particular, if the set of variables $X$ form the entries
of an $n\times n$ matrix, i.e.~$X = \{x_{ij} \mid 1\leq i,j \leq n\}$, then
$\PERM_n$ denotes the polynomial $\sum_{\sigma \in \sym_n}\prod x_{i\sigma(i)}$,
which is the permanent of the matrix.

While a lower bound for the size of general arithmetic circuits computing the
permanent remains out of reach, lower bounds have been established for some
restricted classes of circuits. For example, it is known that there is no
subexponential family of \emph{monotone} circuits for the permanent. This was
first shown for the field of real numbers~\cite{JerrumS82} and a proof for
general fields, with a suitably adapted notion of monotonicity is given
in~\cite{KayalS14}. An exponential lower bound for the permanent is also known
for \emph{depth-3} arithmetic circuits~\cite{GK98} over all finite fields. In
both these cases, the exponential lower bound obtained for the permanent also
applies to the determinant, i.e.~the family of polynomials $\{\DET_n\}_{n \in
  \nats}$, where $\DET_n$ is $\sum_{\sigma \in \sym_n} \sgn{\sigma}\prod
x_{i\sigma(i)}$. However, the determinant is in $\VP$ and so there do exist
polynomial-size families of general circuits for the determinant.

In this paper, we consider a new restriction on arithmetic circuits based on a
natural notion of symmetry, and we show that it distinguishes between the
determinant and the permanent. That is to say, we are able to show exponential
lower bounds on the size of any family of symmetric arithmetic circuits for
computing the permanent, while establishing the existence of polynomial-size
symmetric circuits for computing the determinant.





We next define (informally) the notion of symmetry we use. A formal definition
follows in Section~\ref{sec:symm-circ}.  The permanent and the determinant are
not symmetric polynomials in the usual meaning of the word, in that they are not
invariant under arbitrary permutations of their variables.  However, they do have
natural symmetries, e.g.~permutations of the variables induced by row and column
permutations.  Specifically, $\PERM_n$ is invariant under arbitrary permutations
of the rows and columns of the matrix $(x_{ij})$, while $\DET_n$ is invariant
under a more restricted group of permutations that includes \emph{simultaneous} permutations of the rows and columns.  We consider
similar notions of symmetry on circuits.  We say that an arithmetic circuit $C$
(seen as a labelled directed acyclic graph) that takes as input an $n \times n$
matrix of variables (i.e.\ has input gates labelled by $x_{ij}$, for $i, j \in
[n]$) is \emph{matrix symmetric} if the natural action of any $(\sigma,\pi) \in
\sym(n) \times\sym(n)$ on the inputs (i.e.\ taking $x_{ij}$ to
$x_{\sigma(i)\pi(j)}$) extends to an automorphism of $C$. Similarly, we say $C$
is \emph{square symmetric} if the natural action of any $\sigma \in \sym(n)$ on
its inputs (i.e.\ taking $x_{ij}$ to $x_{\sigma(i)\sigma(j)}$) extends to an
automorphism of $C$.

Our upper bound for the determinant is established for square symmetric circuits over fields of characteristic $0$, and we conjecture it holds for all characteristics. For the permanent we prove exponential lower bounds for square symmetric circuits over fields of characteristic $0$ and for matrix symmetric circuits over all fields of characteristic other than two.  On fields of characteristic two, of course, the permanent and the determinant coincide.

A similar notion of symmetry has been studied previously in the context of
Boolean circuits for deciding graph properties, or properties of relational
structures (see~\cite{DenenbergGS86,Otto96,AndersonD17}). Specifically, such
symmetric circuits arise naturally in the translation into circuit form of
specifications of properties in a logic or similar high-level formalism.
Similarly, we can think of a symmetric arithmetic circuit as a straight-line
program which treats the rows and columns of a matrix as being indexed by
unordered sets. Many natural algorithms have this property. For example, Ryser's
formula for computing the permanent naturally yields a symmetric circuit.

Polynomial-size families of symmetric Boolean circuits with threshold gates form
a particularly robust class, with links to fixed-point
logics~\cite{AndersonD17}. In particular, this allows us to deploy methods for
proving inexpressiblity in such logics to prove lower bounds on the size of
symmetric circuits. A close link has also been established between the power of
such circuits and linear programming extended formulations with a geometric
notion of symmetry~\cite{AtseriasDO19}. Our lower bound for the permanent is
established by first giving a symmetry-preserving translation of arithmetic
circuits to Boolean circuits with threshold gates, and then establishing a lower
bound there for computing the permanent of a $0$-$1$-matrix.

The lower bounds for symmetric Boolean circuits are based on a measure we call
the \emph{counting width} of graph parameters (the term is introduced
in~\cite{DawarW17}). This is also sometimes known as the Weisfeiler-Leman
dimension. In short, we have, for each $k$ an equivalence relation $\equiv^k$,
known as the $k$-dimensional Weisfeiler-Leman equivalence, that is a coarse
approximation of isomorphism, getting finer with increasing $k$. The counting
width of a graph parameter $\mu$ is the smallest $k$, as a function of the graph
size $n$, such that $\mu$ is constant on $\equiv^k$-classes of graphs of size
$n$. From known results relating Boolean circuits and counting
width~\cite{AndersonD17,AtseriasDO19}, we know that the existence of
subexponential size square symmetric circuits computing $\mu$ implies a sublinear upper
bound on its counting width.  Hence, using the relationship between the
permanent of the adjacency matrix of a graph $\Gamma$ and the number of
perfect matchings in $\Gamma$, we obtain our lower bound for the permanent
for square symmetric circuits over fields of characteristic zero by showing a
linear lower bound on the counting width of $\mu(\Gamma)$---the number of perfect
matchings in $G$. Indeed, showing the same for $(\mu(\Gamma) \bmod p)$ for every
prime $p > 2$ allows us to obtain an exponential lower bound for matrix symmetric circuits over any field of characteristic other than two.

The linear lower bound on the counting width of the number of perfect matchings
is a result of interest in its own right, quite apart from the lower bounds it
yields for circuits for the permanent. Indeed, there is an interest in
determining the counting width of concrete graph parameters (see, for
instance,~\cite{ArvindFKV19}), and the result here is somewhat surprising. The
decision problem of determining whether a graph has any perfect matching is
known to have constant counting width. Indeed, the width is $2$ for bipartite
graphs~\cite{BlassGS02}. For general graphs, it is known to be strictly greater
than $2$ but still bounded above by a constant~\cite{AndersonDH15}.

\paragraph*{Related Work.}
While symmetric arithmetic circuits have not previously been studied, symmetric
Boolean circuits have~\cite{DenenbergGS86,Otto96,AndersonD17,Rossman19}.  We review some connections of this previous work with ours in Section~\ref{sec:symm-circ}.  The power of symmetric circuits in the context of computing fully symmetric polynomials can be established from a recent result of Bl\"{a}ser and Jindal, which we review further in Section~\ref{sec:polynomials}.

Landsberg and Ressayre~\cite{LandsbergR16}
establish an exponential lower bound on the complexity of the permanent
(specifically over the complex field $\complex$) under an assumption of
symmetry, and it is instructive to compare our results with theirs. Their lower
bound is for the \emph{equivariant determinantal complexity} of the permanent.  We give a detailed comparison of their results with ours in Section~\ref{sec:equivariant}.  In summary, their result does not yield any lower bounds for symmetric
circuits in the sense we consider.  On the other hand, while we cannot derive their results from ours, but we can derive a lower bound on the determinantal complexity of the permament from our result which is incomparable with theirs.

A preliminary version of this work was presented at the ICALP 2020 conference~\cite{DawarW20}.  The lower bounds we present for the permanent here strengthen the results announced in the conference in two significant ways.  First, we give the lower bound for \emph{square symmetric} circuits, that is, the same symmetry group for which we can prove an upper bound for the determinant.  In the conference paper the lower bound was only given for the stronger condition of matrix symmetric circuits.  Secondly, the lower bound given in the conference was \emph{nearly} exponential: we showed that there is no circuit of size $\mathcal{O}(2^{n^{1-\epsilon}})$ for any positive $\epsilon$.  Here we improve this to show that there is no symmetric circuit for the permanent of size $2^{o(n)}$.

In Section~\ref{sec:background} we introduce some
preliminary definitions and notation. In Section~\ref{sec:symm-circ}, we
introduce the key definitions and properties of symmetric circuits.
Section~\ref{sec:determinant} establishes the upper bound for symmetric circuit
size for the determinant, by translating Le Verrier's method to symmetric
circuits. Finally the lower bounds for the permanent is established in
Sections~\ref{sec:formulas-to-circuits},~\ref{sec:supports},
and~\ref{sec:permanent}. The first of these gives the symmetry-preserving
translation from arithmetic circuits to threshold circuits, the second
establishes an upper-bound on the counting width of a family in terms of the
(orbit) size of the family of circuits deciding it, and the third gives the main
construction proving the linear lower bounds for the counting width of the
number of perfect matchings in a bipartite graph.

%% file: sections-journal/background.tex
In this section we discuss relevant background and introduce notation.

We write $\nats$ for the positive integers and $\natz$ for the non-negative
integers. For $m \in \natz$, $[m]$ denotes the set $\{1, \ldots, m \}$. For a
set $X$ we write $\pow(X)$ to denote the powerset of $X$.

\subsection{Groups}
For a set $X$, $\sym(X)$ is the symmetric group on $X$. For $n \in \nats$ we
write $\sym_n$ to abbreviate $\sym([n])$. The \emph{sign} of a permutation
$\sigma \in \sym(X)$ is defined so that if $\sigma$ is even $\sgn{\sigma} = 1$
and otherwise $\sgn{\sigma} = -1$.

Let $G$ be a group acting on a set $X$. We denote this as a left action, i.e.\
$\sigma x$ for $\sigma \in G$, $x \in X$. The action extends in a natural way to
powers of $X$. So, for $(x,y) \in X \times X$, $\sigma(x,y) = (\sigma x,\sigma
y)$. It also extends to the powerset of $X$ and functions on $X$ as follows. The
action of $G$ on $\pow(X)$ is defined for $\sigma \in G$ and $S \in \pow(X)$ by
$\sigma S = \{\sigma x : x \in S\}$. For $Y$ any set, the action of $G$ on $Y^X$
is defined for $\sigma \in G$ and $f\in Y^X$ by $(\sigma f) (x) = f(\sigma^{-1} x)$
for all $x \in X$. We refer to all of these as the \emph{natural action} of $G$
on the relevant set.

Let $X = \prod_{i \in I} X_i$ and for each $i \in I$ let $G_i$ be a group acting
on $X_i$. The action of the direct product $G := \prod_{i \in I}G_i$ on $X$ is
defined for $x = (x_i)_{i \in I} \in X$ and $\sigma = (\sigma_i)_{i \in I} \in
G$ by $\sigma x = (\sigma_i x_i)_{i \in I}$. If instead $X = \biguplus_{i \in
  I}X_i$ then the action of $G$ on $X$ is defined for $x \in X$ and $\sigma =
(\sigma_i)_{i \in I} \in G$ such that if $x \in X_i$ then $\sigma x = \sigma_i
x$. Again, we refer to either of these as the \emph{natural action} of $G$ on
$X$.

Let $G$ be a group acting on a set $X$. Let $S \subseteq X$. Let $\stab_G(S) :=
\{\sigma \in G : \forall x \in S \,\, \sigma x = x\}$ denote the
\emph{(pointwise) stabiliser} of $S$.

\subsection{Fields and Linear Algebra}

Let $A$ and $B$ be finite non-empty sets. An $A\times B$ \emph{matrix}
with entries in $X$ is a function $M : A \times B \ra X$. For $a \in A$, $b \in
B$ let $M_{ab} = M(a, b)$. We recover the more familiar notion of an $m \times
n$ matrix with rows and columns indexed by ordered sets by taking $A = [m]$ and
$B = [n]$.

The permanent of a matrix is invariant under taking row and column permutations,
while the determinant and trace are invariant under taking \emph{simultaneous}
row and column permutations.  With this observation in mind, we define these
three functions for unordered matrices.  Let $R$ be a commutative ring and $M : A
\times B \ra R$ be a matrix where $ \vert A \vert = \vert B \vert$. Let $\bij(A,
B)$ be the set of bijections from $A$ to $B$. The \emph{permanent} of $M$ over
$R$ is $\perm_R(M) = \sum_{\sigma \in \bij(A, B)}\prod_{a \in A} M_{a
  \sigma(a)}$.  Suppose $A = B$.  The \emph{determinant} of $M$ over $R$ is
$\det_R (M) = \sum_{\sigma \in \sym(A)}\sgn{\sigma}\prod_{a \in A} M_{a
  \sigma(a)}$.  The \emph{trace} of $M$ over $R$ is $\trace_R(M) = \sum_{a \in A}
M_{a a}$.  In all three cases we omit reference to the ring $R$ when it is
obvious from context or otherwise irrelevant.

We always use $\ff$ to denote a field and $\chr(\ff)$ to denote the
characteristic of $\ff$.  For any prime power $q$ we write $\ff_q$ for the
finite field of order $q$.  We are often interested in polynomials defined over a
set of variables $X$ with a natural matrix structure, i.e.\ $X = \{x_{ab} : a
\in A, b \in B\}$.  We identify $X$ with this matrix.  We also identify any
function of the form $f : X \ra Y$ with the $A \times B$ matrix with entries in
$Y$ defined by replacing each $x_{ab}$ with $f(x_{ab})$.

For $n \in \nats$ let $X_n = \{x_{ij} : i, j \in [n]\}$.  Let $\PERM_n :=
\perm(X_n)$ and $\DET_n := \det(X_n)$.  In other words, $\PERM_n$
is the formal polynomial defined by taking the permanent
of an $n \times n$ matrix with $(i, j)$th entry $x_{i j}$, and
similarly for the determinant.
We write $\{\PERM_n\}$ to abbreviate $\{\PERM_n : n \in \nats\}$ and $\{\DET_n\}$
to abbreviate $\{\DET_n : n \in \nats\}$.

\subsection{Graphs, Matrices and Matchings}
Given a graph $\Gamma = (V,E)$, the adjacency matrix $A_G$ of $\Gamma$ is the $V \times V$ $\{0,1\}$-matrix with $A_\Gamma(u,v) = 1$ if, and only if, $\{u,v\} \in E$.  If $\Gamma$ is bipartite, with bipartition $V = A \cup B$, then the \emph{biadjacency matrix} $B_\Gamma$ of $\Gamma$ is the $A \times B$ $\{0,1\}$-matrix with $B_\Gamma(u,v) = 1$ if, and only if, $\{u,v\} \in E$.

It is well known that for a bipartite graph $\Gamma$, $\perm(B_\Gamma)$ over any field of characteristic zero counts the number of perfect matchings in $\Gamma$~\cite{Harary69} and for prime $p$, $\perm_{\ff}(B_\Gamma)$ for a field $\ff$ of characteristic $p$ counts the number of perfect matchings in $\Gamma$ modulo $p$. For bipartite $\Gamma$, 
$A_\Gamma$ is a block anti-diagonal matrix with two blocks corresponding to $B_\Gamma$ and $B_\Gamma^T$ and $\perm(A_\Gamma) = \perm(B_\Gamma)^2$.

\subsection{Counting Width}
For any $k \in \nats$, the $k$-dimensional Weisfeiler-Leman equivalence
(see~\cite{CFI92}), denoted $\equiv^k$ is an equivalence relation on graphs that
provides an over-approximation of isomorphism in the sense that for isomorphic
graphs $\Gamma$ and $\Delta$, we have $\Gamma \equiv^k \Delta$ for all $k$. Increasing values of $k$
give finer relations, so $\Gamma \equiv^{k+1} \Delta$ implies $\Gamma \equiv^k \Delta$ for all $k$.
The equivalence relation is decidable in time $n^{\mathcal{O}(k)}$, where $n$ is
the size of the graphs. If $k \geq n$, then $\Gamma \equiv^k \Delta$ implies that $\Gamma$ and
$\Delta$ are isomorphic. The Weisfeiler-Leman equivalences have been widely studied
and they have many equivalent characterizations in combinatorics, logic, algebra
and linear optimization.   One particularly useful characterization in
terms of logic (see~\cite{CFI92}) is that $\Gamma \equiv^k \Delta$ if,
and only if, $\Gamma$ and $\Delta$ cannot be distinguished by any
formula of first-order logic with \emph{counting quantifiers} using at most $k+1$ distinct variables.  This has been used to  establish
inexpressibility results in various counting logics and motivates the notion of
\emph{counting width}.

A \emph{graph parameter} is a function $\mu$ from graphs to a set $X$ which is
isomorphism invariant.  That is to say, $\mu(\Gamma) = \mu(\Delta)$ whenever $\Gamma$ and $\Delta$ are isomorphic graphs.  Most commonly, $X$ is the set $\nats$ and examples of such graph parameters are the chromatic number, the number of
connected components or the number of perfect matchings.  We can also
let $X$ be a field $\ff$ and let $\mu(\Gamma)$ denote the permanent
(over $\ff$) of the adjacency matrix of $\Gamma$.  When $X = \{0,1\}$
we identify $\mu$ with the class of graphs for which it is the
indicator function.  In this case, we also call it a \emph{graph property}.

For a graph parameter
$\mu$ and any fixed $n \in \nats$, there is a smallest value of $k$ such that
$\mu$ is $\equiv^k$-invariant on graphs with at most $n$ vertices. This motivates the definition.
\begin{definition}\label{def:counting-width}
  For any graph parameter $\mu$, the \emph{counting width} of $\mu$ is the
  function $\nu : \nats \ra \nats$ such that $\nu(n)$
  is the smallest $k$ such that for all graphs $\Gamma, \Delta \in \mathcal{C}$ of size
  at most $n$, if $\Gamma \equiv^{k} \Delta$, then $\mu(\Gamma) = \mu(\Delta)$.
\end{definition}
The notion of counting width for classes of
graphs was introduced in~\cite{DawarW17}, which we here extend to graph
parameters. Note that for any graph parameter $\nu(n) \leq n$, since
any non-isomorphic graphs on $n$ vertices can allways be distinguished
in $\equiv^n$.

Cai, F\"{u}rer and Immerman~\cite{CFI92} first showed that there is no fixed $k$
for which $\equiv^k$ coincides with isomorphism. Indeed, in our terminology,
they construct a graph property with counting width $\Omega(n)$. Since then,
many graph properties have been shown to have linear counting width, including
Hamiltonicity and 3-colourability
(see~\cite{AtseriasDO19}). In other cases, such as the class of graphs that
contain a perfect matching, it has been proved that they have counting width
bounded by a constant~\cite{AndersonDH15}. Our interest in counting width stems
from the relation between this measure and lower bounds for symmetric circuits.
Roughly, if a class of graphs is recognized by a family of
polynomial-sized symmetric threshold circuits, it has bounded counting width (a
more precise statement is given in
Theorem~\ref{thm:cw-graphs}).

Our lower bound construction in Section~\ref{sec:permanent} is based on the
graphs constructed by Cai et al.~\cite{CFI92}. While we review some of the
details of the construction in Section~\ref{sec:permanent}, a reader unfamiliar
with the construction may wish to consult a more detailed introduction. The
original construction can be found in~\cite{CFI92} and a version closer to what
we use is given in~\cite{DR07}.

\subsection{Circuits}
We provide a general definition that incorporates both Boolean and arithmetic
circuits.
\begin{definition}[Circuit]
  A \emph{circuit} over the \emph{basis} $\BB$ with \emph{variables} $X$ and
  \emph{constants} $K$ is a directed acyclic graph with a labelling where each
  vertex of in-degree $0$ is labelled by an element of $X \cup K$ and each
  vertex of in-degree greater than $0$ is labelled by an element of $b \in \BB$ such that the arity of the basis element $b$ matches the in-degree of the gate.
\end{definition}
Note that, in the examples we consider, the elements of the basis
often do not have fixed arity.  That is, we are considering unbounded
fan-in circuits where gates such as AND, OR, $+$, $\times$ can take
any number of inputs.  The one exception is the NOT gate.

Let $C = (D, W)$, where $W \subset D \times D$, be a circuit with constants $K$.
We call the elements of $D$ \emph{gates}, and the elements of $W$ \emph{wires}.
We call the gates with in-degree $0$ \emph{input gates} and gates with
out-degree $0$ \emph{output gates}. We call those input gates labelled by
elements of $K$ \emph{constant gates}. We call those gates that are not input
gates \emph{internal gates}. For $g, h \in D$ we say that $h$ is a \emph{child}
of $g$ if $(h, g) \in W$. We write $\child{g}$ to denote the set of children of
$g$. We write $C_g$ to denote the sub-circuit of $C$ rooted at $g$. Unless
otherwise stated we always assume a circuit has exactly one output gate. We also assume that distinct input gates in a circuit have distinct labels.

If $K$ is a field $\ff$, and $\BB$ is the set $\{+, \times\}$, we have an
\emph{arithmetic circuit} over $\ff$. If $K = \{0,1\}$, and $\BB$ is a
collection of Boolean functions, we have a \emph{Boolean circuit} over the basis
$\BB$. We define two Boolean bases here. The \emph{standard basis} $\BS$
contains the functions $\land$, $\lor$, and $\neg$. The \emph{threshold basis}
$\BT$ is the union of $\BS$ and $\{t_{\geq k} : k \in \nats\}$, where for each
$k \in \nats$, $t_{\geq k}$ is defined for a string $\vec{x} \in \{0, 1\}^*$ so
that $t_{\geq k}(\vec{x}) = 1$ if, and only if, the number of $1$s in $\vec{x}$
is at least $k$. We call a circuit defined over this basis a \emph{threshold
  circuit}. Another useful function is $t_{=k}$, which is defined by $t_{=k}(x)
= t_{\geq k}(x) \land \neg t_{\geq k+1}(x)$. We do not explicitly include it in
the basis as it is easily defined in $\BT$.

In general, we require that a basis contain only functions that are invariant
under all permutations of their inputs (we define this notion formally in
Definition~\ref{def:full-symm}). This is the case for the arithmetic functions
$+$ and $\times$ and for all of the Boolean functions in $\BT$ and $\BS$. Let
$C$ be a circuit defined over such a basis with variables $X$ and constants $K$.
We evaluate $C$ for an assignment $M \in K^X$ by evaluating each gate labelled
by some $x \in X$ to $M(x)$ and each gate labelled by some $k \in K$ to $k$, and
then recursively evaluating each gate according to its corresponding basis
element. We write $C[M](g)$ to denote the value of the gate $g$ and $C[M]$ to
denote the value of the output gate. We say that $C$ computes the function $M
\mapsto C[M]$.

It is conventional to consider an arithmetic circuit $C$ over $\ff$ with
variables $X$ to be computing a polynomial in $\ff[X]$, rather than a function
$\ff^X \ra \ff$. This polynomial is defined via a similar recursive evaluation,
except that now each gate labelled by a variable evaluates to the corresponding
formal variable, and we treat addition and multiplication as ring operations in
$\ff[X]$. Each gate then evaluates to some polynomial in $\ff[X]$. The
polynomial computed by $C$ is the value of the output gate.

For more details on arithmetic circuits see~\cite{ShpilkaY10} and for Boolean
circuits see~\cite{Vollmer99}.

  By a standard translation (see~\cite{Valient1981}),  arithmetic circuits with unbounded
  fan-in can be mapped to equivalent arithmetic circuits with constant fan-in with only a polynomial
  blowup in size and a logarithmic blowup in depth. This means that so long as we are interested in bounds on circuit size up to polynomial factors we may assume without a loss
  of generality that all gates have fan-in two. This assumption simplifies the
  analysis of these circuits and in many cases authors simply define arithmetic
  circuits to have internal gates with fan-in two (e.g.~\cite{ShpilkaY10}). In
  this paper we are interested in symmetric arithmetic circuits and
  the standard
  translation does not preserve symmetry. As such, we cannot assume a bound on
  fan-in without a loss of generality and for this reason we define arithmetic circuits so as to
  allow for unbounded fan-in.

%% file: sections-journal/symmetric-arithmetic-circuits.tex
In this section we discuss different symmetry conditions for functions and
polynomials. We also introduce the notion of a symmetric circuit.

\subsection{Symmetric Functions}\label{sec:symmetric-functions}

\begin{definition}
  For any group $G$, we say that a function $F: K^X \ra K$, along with an action
  of $G$ on $X$ is a \emph{$G$-symmetric function}, if for every $\sigma \in G$,
  $\sigma F = F$.
\end{definition}

We are interested in some specific group actions, which we now define and
illustrate with examples.

\begin{definition}
  \label{def:full-symm}
  If $G=\sym(X)$, we call a $G$-symmetric function $F: K^X \ra K$, \emph{fully
    symmetric}.
\end{definition}

Examples of fully symmetric functions are those that appear as labels of gates
in a circuit, including $+$, $\times$, $\land$, $\lor$ and $t_{\geq k}$.

\begin{definition}
  If $G = \sym(X) \times \sym(Y)$ and $F : K^{X \times Y} \ra K$ is
  $G$-symmetric with the natural action of $G$ on $X \times Y$, then we say $F$
  is \emph{matrix symmetric}.
\end{definition}

Matrix symmetric functions are those where the input is naturally seen as a
matrix with the result invariant under aribtrary row and column permutations.
The canonical example for us of a matrix symmetric function is the permanent.
The determinant is not matrix symmetric over fields of characteristic other than
$2$, but does satisfy a more restricted notion of symmetry that we define next.


\begin{definition}
  If $G = \sym(X)$ and $F : K^{X \times X} \ra K$ is $G$-symmetric with the
  natural action of $G$ on $X \times X$, then we say $F$ is \emph{square
    symmetric}.
\end{definition}

The determinant is one example of a square symmetric function. However, as the
determinant of a matrix is also invariant under the operation of transposing the
matrix, we also consider this variation. To be precise, let $\sigma_t \in \sym(X
\times X)$ be the permutation that takes $(x,y)$ to $(y,x)$ for all $x,y\in X$.
Let $G_{\sqr}$ be the diagonal of $\sym(X)\times \sym(X)$ (i.e.\ the image of
$\sym(X)$ in its natural action on $X \times X$). We write $G_{\transp}$ for the
group generated by $G_{\sqr} \cup \{\sigma_t\}$. We say a
$G_{\transp}$-symmetric function is \emph{transpose symmetric}.





Finally, another useful notion of symmetry in functions is where the inputs are
naturally partitioned into sets.

\begin{definition}
  If $X = \biguplus_{i \in I} X_i$, $G = \prod_{i \in I}\sym(X_i)$, and $F: K^X
  \ra K$ is $G$-symmetric with respect to the natural action of $G$ on $X$, we
  say $F$ is \emph{partition symmetric}.
\end{definition}

In Section~\ref{sec:formulas-to-circuits}, we consider a generalization of
circuits to the case where the labels in the basis are not necessarily fully
symmetric functions, but they are still partition symmetric. The structure of
such a circuit can not be described simply as a DAG, but requires additional
labels on wires, as we shall see.





\subsection{Symmetric Circuits}
Symmetric Boolean circuits have been considered in the literature, particularly
in connection with definability in logic. In that context, we are considering
circuits which take relational structures (such as graphs) as inputs and we
require their computations to be invariant under re-orderings of the elements of
the structure.  Thus, the inputs to a Boolean circuit $C$ might be labelled by pairs of elements $(x,y)$ where $x,y \in V$ and we require the output of $C$ to be invariant under a permutation of $V$ applied to the inputs.  In short, the function computed by $C$ is square symmetric.  A generalization to arbitrary symmetry groups was also defined by Rossman~\cite{Rossman19} who showed a lower bound for the parity function for formulas that are $G$-symmetric for subgroups $G$ of $\ZZ_2^n$.
Here, we consider circuits that are symmetric with respect to arbitrary symmetry groups, and also consider them in the context of arithmetic circuits. In order to define
symmetric circuits, we first need to define the automorphisms of a circuit.

\begin{definition}[Circuit Automorphism]
  Let $C = (D, W)$ be a circuit over the basis $\BB$ with variables $X$ and
  constants $K$. For $\sigma \in \sym(X)$, we say that a bijection $\pi : D \ra
  D$ is an \emph{automorphism} extending $\sigma$ if for every gate $g$ in $C$
  we have that
  \begin{itemize}
  \item if $g$ is a constant gate then $\pi (g) = g$,
  \item if $g$ is a non-constant input gate then $\pi (g) = \sigma (g)$,
  \item if $(h,g) \in W$ is a wire, then so is $(\pi h, \pi g)$
  \item if $g$ is labelled by $b \in \BB$, then so is $\pi g$.
  \end{itemize}
\end{definition}

We say that a circuit $C$ with variables $X$ is \emph{rigid} if for every
permutation $\sigma \in \sym(X)$ there is at most one automorphism of $C$
extending $\sigma$.

We are now ready to define the key notion of a symmetric circuit.
\begin{definition}[Symmetric Circuit]
  Let $C$ be a circuit with variables $X$ and $G \leq \sym_X$. We say $C$ is
  \emph{$G$-symmetric} if the action of every $\sigma \in G$ on $X$ extends to
  an automorphism of $C$. We say that $C$ is \emph{strictly $G$-symmetric} if the only automorphisms of $C$ are those extending a permutation in $G$.
\end{definition}

It is easy see that if a circuit is $G$-symmetric then it computes a
$G$-symmetric polynomial (and hence function). We sometimes omit mention of $G$
when it is obvious from context. For a gate $g$ in a symmetric circuit $C$, the
\emph{orbit} of $g$, denoted by $\orb(g)$, is the the set of all $h \in C$ such
that there exists an automorphism $\pi$ of $C$ extending some permutation in $G$ with $\pi(g) = h$. We write
$\ORB{C}$ for the maximum size of an orbit in $C$, and call it the \emph{orbit
  size} of $C$.

We use the same terminology for symmetric circuits as for symmetric functions.
That is, if a circuit $C$ with variables $X \times Y$ is $\sym_X \times
\sym_Y$-symmetric we say that $C$ is matrix symmetric. We similarly define
square symmetric circuits, transpose symmetric circuits and partition symmetric circuits. 

Though symmetric arithmetic circuits have not previously been studied, symmetric
Boolean circuits have~\cite{DenenbergGS86,Otto96,AndersonD17,Rossman19}. It is known that
polynomial-size square symmetric threshold circuits are more powerful than
polynomial-size square symmetric circuits over the standard basis~\cite{AndersonD17}.
In particular, the majority function is not computable by any family of
polynomial-size symmetric circuits over the standard basis. On the other hand,
it is also known~\cite{DawarW22} that adding any fully symmetric functions to
the basis does not take us beyond the power of the threshold basis. Thus, $\BT$
gives the robust notion, and that is what we use here. It is also this that has
the tight connection with counting width mentioned above.




\subsection{Polynomials}\label{sec:polynomials}
In the study of arithmetic complexity, we usually think of a circuit over a
field $\ff$ with variables in $X$ as expressing a polynomial in $\ff[X]$, rather
than computing a function from $\ff^X$ to $\ff$. The distinction is signficant,
particularly when $\ff$ is a finite field, as it is possible for distinct
polynomials to represent the same function.

The definitions of symmetric functions given in
Section~\ref{sec:symmetric-functions} extend easily to polynomials. So, for a
group $G$ acting on $X$, a polynomial $p \in \ff[X]$ is said to be $G$-symmetric
if $\sigma p = p$ for all $\sigma \in G$.   It is clear that a $G$-symmetric polynomial
determines a $G$-symmetric function.  We define \emph{fully symmetric},
\emph{matrix symmetric}, \emph{square symmetric} and \emph{transpose symmetric}
polynomials analogously. Every matrix symmetric polynomial is also square
symmetric. Also, every transpose symmetric polynomial is square symmetric. The
permanent $\PERM_n$ is both matrix symmetric and transpose symmetric, while the
determinant $\DET_n$ is transpose symmetric, but not matrix symmetric.

What are usually called the \emph{symmetric polynomials} are, in our
terminology, fully symmetric. In particular, the homogeneous polynomial $\sum_{i
  \in [n]}x_i^r$ is fully symmetric. There is a known lower bound of $\Omega(n
\log r)$ on the size of any circuit expressing this polynomial~\cite{BaurS83}.
It is worth remarking that the matching upper bound is achieved by a fully symmetric
circuit. Thus, at least in this case, there is no gain to be made by breaking
symmetries in the circuit.  Similarly, we have tight quadratic upper and lower
bounds for depth-3 circuits for the elementary symmetric polynomials $\sum_{S \subseteq [n] : |S|=k}\prod_{i\in
  S}x_i$ over infinite fields~\cite{ShpilkaW01}.  The upper
bound is obtained  by the interpolation method and it can be seen that this is achieved by
fully symmetric circuits.  To be precise, the polynomial is computed
as the coefficient of $t^{n-k}$ in $\prod_{i=1}^n (t+x_i)$, which is
obtained by interpolation from computing $\prod_{i=1}^n (t+x_i)$ at
$n+1$ distinct values of $t$.  Note that, for any fixed constant $t$,
$\prod_{i=1}^n (t+x_i)$ is given by a fully symmetric circuit of size
$O(n)$, and these can be combined to get the interpolant.  The
resulting circuit is still fully symmetric since a permutation of the
variables $x_i$ fixes the polynomial $\prod_{i=1}^n (t+x_i)$.

Indeed, we can say something more general about fully symmetric
polynomials.  If any such polynomial $f\in \mathbb{C}[x_1,...,x_n]$ has a circuit of
size polynomial in $n$ then it has a $\sym_n$-circuit of size polynomial in $n$. This follows from a result of Bl\"{a}ser and
Jindal~\cite{BlaeserJ19} who establish that for any fully symmetric
polynomial $f \in \mathbb{C}[x_1, \ldots, x_n]$ which has a
polynomial-size circuit there exists a witness
$g\in \mathbb{C}[y_1, \ldots, y_n]$ computable via an arithmetic
circuit of size polynomial in $n$ such that $f = g(e_1, \ldots, e_n)$,
where the $e_i$'s are the elementary symmetric polynomials.   To see
why this implies the result, observe that if $f$ is a fully symmetric
polynomial and $g$ is the corresponding witness computable via a
polynomial size circuit $C$, and $E_i$ are the (fully symmetric and
polynomial size) circuits computing the polynomials $e_i$, then we can
build a circuit for $f$ by replacing each input $y_i$ in $C$ with the
output gate of $E_i$. The resultant circuit is symmetric since any
permutation on the input gates fixes the output gate of each $E_i$. 
 
The best known upper bound for general arithmetic circuits for expressing the
permanent is given by Ryser's formula:
$$\PERM_n  = (-1)^n \sum_{S \subseteq [n]} (-1)^{|S|} \prod_{i=1}^n
\sum_{j\in S} x_{ij}.$$ It is easily seen that this expression is matrix symmetric, and
it yields a matrix symmetric circuit of size $\mathcal{O}(2^{n}n^2)$.  Our main result,
Theorem~\ref{thm:lowerbound-square}, gives us a near matching lower bound on the size of matrix symmetric circuits (or even square symmetric circuits) for expressing $\PERM_n$.

A $G$-symmetric circuit $C$ expressing a polynomial $p$ is also a
$G$-symmetric circuit computing the function determined by $p$. In establishing our
upper bound for the determinant, we show the existence of small transpose symmetric
circuits for the polynomial, and hence also for the function. For the lower
bound on the permanent, we show that there are no small square symmetric circuits for
computing the function, hence also none for the polynomial. For a discussion of
functional lower bounds, as opposed to polynomial lower bounds,
see~\cite{ForbesKS16}.

%% file: sections-journal/determinant.tex
In this section we show that for any field $\ff$ with characteristic zero there
is a polynomial-size family of transpose symmetric arithmetic circuits over $\ff$
computing $\{\DET_n\}$. We define this family using Le Verrier's method for
calculating the characteristic polynomial of a matrix. We review this method
briefly, and direct the reader to Section~3.4.1 in~\cite{Holm10} for more
detail.

The characteristic polynomial of an $n \times n$ matrix $M$ is
\begin{align*}
  \det(x I_n - M) = \prod^n_{i = 1} (x - \lambda_i) = x^n - p_1x^{n-1} + p_2 x^{n -2} - \ldots + (-1)^{n}p_n,
\end{align*}
where $\lambda_1, \ldots , \lambda_n$ are the eigenvalues of $M$, counted with
multiplicity. It is known that $p_n = \det (M)$ and $p_1 = \trace(M)$. Le
Verrier's method gives, for each $i \in [n]$, the linear recurrence given by
$$p_i = \frac{1}{i} [p_{i -1} s_1 - p_{i -2} s_2 + \ldots \pm s_i],$$
where $p_0 = 1$ and for each $j \in [n]$, $s_j = \trace(M^j)$.

The determinant can thus be computed as follows. First, for each $k \in [n]$ we
compute entries in the matrix $M^k$. Then for each $k \in [n]$ we compute $s_k =
\trace(M^k)$. Finally, we recursively compute each $p_i$ and output $p_n$. There
is a natural arithmetic circuit $\Phi$ with variables $M = \{m_{ij} : i, j \in
[n]\}$ implementing this algorithm.

To see that $\Phi$ is transpose symmetric we begin with some permutation $\sigma
\in G_{\transp}$ and show that $\sigma$ extends to an automorphism of the
circuit. We construct this automorphism layer by layer. We map each gate
computing some $(i, j)$ entry of $M^k$ to the gate computing the $\sigma (i, j)$
entry of $M^k$. We fix each gate computing some $s_k$. Since each gate computing
some $p_l$ uses only the gates computing $s_1, \ldots, s_k$ and a constant gate
computing $1$, we can also fix each of these gates. We now present this argument
formally.

\begin{theorem}
  For $\ff$ be a field of characteristic $0$, there exists a family of transpose
  symmetric arithmetic circuits $(\Phi_n)_{n \in \nats}$ over $\ff$ computing
  $\{\DET_n\}$ for which the function $n \mapsto \Phi_n$ is computable in time
  $\mathcal{O}(n^4)$.
  \label{thm:det-upper-bound}
\end{theorem}
\begin{proof}
  Let $n \in \nats$ and let $X = (x_{ij})_{i, j \in I}$ be an $I \times I$
  matrix of variables, for an index set $I$ with $\vert I \vert=n$. We now
  describe an implementation of Le Verrier's method for $I \times I$ matrices as
  arithmetic circuit $\Phi_n$ over the set of variables $X$. We construct this
  circuit as follows.
  \begin{itemize}
  \item For each $k \in [n]$ we include a family of gates intended to compute
    the entries in the $k$th power of the matrix $X$. For each $i, j \in I$ we
    include a gate $(k; i, j)$ intended to compute $(X^k)_{ij}$. Let
    $(\Phi_n)_{(2; i ,j)} = \sum_{a \in I} x_{ia} x_{aj}$ and for all $k > 2$,
    $(\Phi_n)_{(k; i ,j)} = \sum_{a \in I} (\Phi_n)_{(k-1; i, a)} x_{a, j}$.
  \item For each $k \in [n]$ we include a gate $(\trace, k)$ intended to compute
    the trace of $X^k$. Let $(\Phi_n)_{(\trace, 1)} = \sum_{a \in I} x_{a, a}$
    and for $k > 1$, $(\Phi_n)_{(\trace, k)} = \sum_{a \in I} (\Phi_n)_{(k; a,
      a)}$.
  \item For each $k \in [n]$ we include a gate $(p, k)$ intended to compute the
    coefficient $p_k$ in the characteristic polynomial. Let $(\Phi_n)_{(p, 1)} =
    (\Phi_n)_{(\trace, 1)}$ and for all $k > 1$ let
    \begin{align*}
      (\Phi_n)_{(p, k)} = \frac{1}{k}[&(\Phi_n)_{(p, k - 1)} (\Phi_n)_{(\trace, 1)} - (\Phi_n)_{(p, k - 2)} (\Phi_n)_{(\trace, 2)} \\ &+ (\Phi_n)_{(p, k - 3)} (\Phi_n)_{(\trace, 3)} - \ldots \pm (\Phi_n)_{(\trace, k)}].
    \end{align*}
  \end{itemize}
  Let $(p, n)$ be the output gate of $\Phi_n$. It follows from the discussion
  preceding the statement of the theorem that $(p, n)$ computes $\DET_n$.

  It remains to show that the circuit is transpose symmetric. Let $\sigma \in
  \sym(I)$. Let $\pi : \Phi_n \ra \Phi_n$ be defined such that for each input
  gate labelled $x_{ij}$ we have $\pi (x_{ij}) = x_{\sigma(i)\sigma(j)}$, for
  each gate of the form $(k; i, j)$ we have $\pi (k; i, j) = (k;
  \sigma(i),\sigma(j))$, and for every other gate $g$ we have $\pi (g) = g$. It
  can be verified that $\pi$ is a circuit automorphism extending $\sigma$.
  Similarly, if $\sigma_t \in \sym(I \times I)$ is the transpose permutation,
  i.e.~$\sigma_t(i,j) = (j,i)$, then we can extend it to an automorphism $\pi_t$
  of $\Phi_n$ by letting $\pi_t (k;i,j) = (k;j,i)$. It follows that $\Phi_n$ is
  a transpose symmetric arithmetic circuit.

  The circuit contains constant gates labelled by $-1, 0, 1, \frac{1}{2} \ldots
  \frac{1}{n}$.  There are $n^2$ other input gates. Computing each gate $(k; i, j)$ uses $O(1+n)$ gates ($n$ products and $1$ sum). Then, since there are $n^2$ entries in each matrix and $n-1$ matrices to compute, the total number of gates needed to compute all of the $(k;i,j)$ gates is $O(n^2 (n-1) (n+1))$. There are $n$ additional gates required to compute all gates of the form $(\trace,
  i)$.  There are at most $n (2n -1)$  gates required to compute all
  gates of the form $(p, i)$. It follows that the circuit is of size
  $\mathcal{O}(n^4)$. The above description of the circuit $\Phi_n$ can be
  adapted to define an algorithm that computes the function $n \mapsto \Phi_n$
  in time $\mathcal{O}(n^4)$.
\end{proof}

Le Verrier's method explicitly involves multiplications by field elements
$\frac{1}{k}$ for $k \in [n]$, and so cannot be directly applied to fields of
positive characteristic. We conjecture that it is also possible to give square
symmetric arithmetic circuits of polynomial size to compute the determinant over
arbitrary fields. Indeed, there are many known algorithms that yield
polynomial-size families of arithmetic circuits over fields of positive
characteristic computing $\{\DET_n\}$. It seems likely that some of these could
be implemented symmetrically.

%% file: sections-journal/arithmetic-boolean.tex
In this section we establish the following symmetry and orbit-size preserving translation from
arithmetic circuits to threshold circuits. Importantly, this translation does not preserve circuit size, which may grow exponentially.

\begin{theorem}
  Let $G$ be a group acting on a set of variables $X$. Let $\Phi$ be a
  $G$-symmetric arithmetic circuit over a field $\ff$ with variables $X$. Let $B
  \subseteq \ff$ be finite. Then there is a $G$-symmetric threshold circuit $C$
  with variables $X$ and $\ORB{C} = \ORB{\Phi}$, such that for all $M \in \{0, 1\}^X$ we have $C[M] = 1$
  if, and only if, $\Phi[M] \in B$.
  \label{thm:arithmetic-to-boolean}
\end{theorem}

We use Theorem~\ref{thm:arithmetic-to-boolean} in Section~\ref{sec:permanent} to
transfer a lower bound on threshold circuits to arithmetic circuits, a crucial
step in establishing our lower bound for the permanent. This lower bound relies on the preservation of orbit-size in Theorem~\ref{thm:arithmetic-to-boolean}, and the connection between orbit-size and counting width.

We prove Theorem~\ref{thm:arithmetic-to-boolean} by first establishing a similar
translation from arithmetic circuits over a field $\ff$ to Boolean circuits over
a basis $\BA^\ff$ of partition symmetric functions. We then complete the proof
by replacing each gate labelled by a partition symmetric function with an
appropriate symmetric Boolean threshold circuit.

To enable this second step, we first show that each partition symmetric function
can be computed by a rigid strictly symmetric threshold circuit. The proof of
this follows from the fact that if a function $F: \{0, 1\}^A \ra \{0, 1\}$ is
partition symmetric, then its output for $h \in \{0, 1\}^A$ depends only on the
\emph{number} of elements in each part of $A$ that $h$ maps to $1$. We can thus
evaluate $F$ by counting the number of $1$s in each part, a procedure which we
now show can be implemented via a symmetric threshold circuit.

\begin{lemma}
  Let $F$ be a partition symmetric function. There exists a rigid strictly
  partition symmetric threshold circuit $C(F)$ computing $F$.
  \label{lem:partition-symmetric}
\end{lemma}
\begin{proof}
  Let $A := \biguplus_{q \in Q}A_q$ be a disjoint union of finite sets $A_q$
  indexed by $Q$, and $F : \{0, 1\}^A \ra \{0, 1\}$ be a partition symmetric
  function. The fact that $F$ is partition symmetric means that whether $F(h) =
  1$ for some $h \in \{0,1\}^A$ is determined by the number of $a \in A_q$ (for
  each $q$) for which $h(a) = 1$. Write $h_q$ for this number. Then, there is a
  set $c_F \subseteq \natz^Q$ such that $F(h) = 1$ if, and only if, $(h_q)_{q\in
    Q} \in c_F$. Since each $A_q$ is finite, so is $c_F$. Then $F(h) = 1$ if,
  and only if, the following Boolean expression is true: $\bigvee_{c \in
    c_F}\bigwedge_{q\in Q} (h_q = c(q)).$ We can turn this expression into a
  circuit $C$ with an $\OR$ gate at the output, whose children are $\AND$ gates,
  one for each $c \in c_F$, let us call it $\land_c$. The children of $\land_c$
  are a set of gates, one for each $q \in Q$, let us call it $T_{c,q}$, which is
  labelled by $t_{=c(q)}$ and has as children all the inputs $a \in A_q$.

  This circuit $C$ is symmetric and rigid, but not necessarily strictly
  symmetric, as it may admit automorphisms that do not respect the partition of
  the inputs $A$ as $\biguplus_{q \in Q}A_q$. To remedy this, we create pairwise
  non-isomorphic gadgets $D_q$, one for each $q \in Q$. Each $D_q$ is a
  one-input, one-output circuit computing the identity function. For example,
  $D_q$ could be a tower of single-input $\AND$ gates, and we choose a different
  height for each $q$. We now modify $C$ to obtain $C(F)$ by inserting between
  each input $a \in A_q$ and each gate $T_{c,q}$ a copy $D^a_q$ of the gadget
  $D_q$.

  Clearly $C(F)$ computes $F$. We now argue $C(F)$ is rigid and strictly partition
  symmetric. To see that it is partition symmetric, consider any $\sigma \in \prod_{q \in
    Q}\sym(A_q)$ in its natural action on $A$. This extends to an automorphism
  of $C(F)$ that takes the gadget $D^a_q$ to $D^{\sigma a}_q$ while fixing all
  gates $T_{c,q}$ and $\land_c$. To see that there are no other automorphisms,
  suppose $\pi$ is an automorphism of $C(F)$. It must fix the output $\OR$ gate.
  Also $\pi$ cannot map a gate $T_{c,q}$ to $T_{c',q'}$ for $q'\neq q$ because
  the gadgets $D_q$ and $D_{q'}$ are non-isomorphic. Suppose that $\pi$ maps
  $\land_c$ to $\land_{c'}$. Then, it must map $T_{c,q}$ to $T_{c',q}$. Since
  the labels of these gates are $t_{=c(q)}$ and $t_{=c'(q)}$ respectively, we
  conclude that $c(q) = c'(q)$ for all $q$ and therefore $c=c'$.
\end{proof}

We now define for each field $\ff$ the basis $\BA^{\ff}$. The functions in this
basis are intended to be Boolean analogues of addition and multiplication. Let
$Q \subseteq \ff$ be finite, $A = \biguplus_{q \in Q}A_q$ be a disjoint union of
non-empty finite sets, and $c \in \ff$.  Formally, we define for any $h \in \{0,1\}^{A}$ the functions $+^A_{Q, c}: \{0,
1\}^A \ra \{0, 1\}$ and $\times^A_{Q, c}: \{0, 1\}^A \ra \{0, 1\}$ as follows: $ +^{A}_{Q, c}(h) = 1
\text{ if, and only if,} \sum_{q \in Q} \vert \{a \in A_q : h(a) = 1\} \vert
\cdot q = c $ and $ \times^{A}_{Q, c}(h) = 1 \text{ if, and only if, } \prod_{q
  \in Q} q^{\vert \{a \in A_q : h(a) = 1\} \vert} = c$. Both $+^A_{Q, c}$ and
$\times^A_{Q, c}$ are partition symmetric. Let $\BA^\ff$ be the set of all
functions $+^A_{Q, c}$ and $\times^A_{Q, c}$.

We aim to prove Theorem~\ref{thm:arithmetic-to-boolean} by first defining for a
given $G$-symmetric arithmetic circuit a corresponding $G$-symmetric Boolean
circuit over a partition symmetric basis. To ensure unambiguous evaluation, the
circuit must include for each gate labelled by a partition symmetric function a
corresponding partition on its children. Let $C$ be a circuit with variables $X$
and let $g$ be a gate in $C$ labelled by a partition symmetric function $F :
\{0, 1\}^A \ra \{0, 1\}$, where $A = \biguplus_{q \in Q}A_q$ is a disjoint union
of finite non-empty sets. We associate with $g$ a bijection $L_g : A \ra
\child{g}$. We evaluate $g$ for an input as follows. For $M \in \{0, 1\}^{X}$ we
let $L^M_g : A \ra \{0, 1\}$ be defined such that $L^M_g(a) = C[M](L_g(a))$ for
all $a \in A$. Let $C[M](g) = F(L^M_g)$.

\begin{proof}[Proof of Theorem~\ref{thm:arithmetic-to-boolean}]
  We associate with each $v \in \Phi$ a finite set $Q_v \subseteq \ff$
  such that for any assignment of $0$-$1$ values to the inputs, $M \in
  \{0,1\}^X$, we have $\Phi[M](v) \in Q_v$.  This can be defined by
  induction on the structure of $\Phi$: If $v$ is an input gate, $Q_v
  = \{0,1\}$; and if $v$ is an $\odot$-gate for $\odot \in
  \{+,\times\}$ with children $u_1,\ldots,u_t$  we let $Q_v = \{
  a_1\odot \cdots \odot a_t \mid a_i \in Q_{u_i} \}$.
  Let $z$ be the output gate of $\Phi$. If $Q_z \subseteq B$ let $C$ be the
  circuit consisting of a single gate labelled by $1$ and if $Q_z \cap B =
  \emptyset$ let $C$ consist of a single gate labelled by $0$. Suppose that
  neither of these two cases hold.

  We now construct a $\BA^{\ff}\cup \BS$-circuit $D$ from $\Phi$ by replacing
  each internal gate $v$ in $\Phi$ with a family of gates $(v, q)$ for $q \in
  Q_v$ such that $D[M](v, q) = 1$ if, and only if, $\Phi[M](v) = q$. Each $(v,
  q)$ is labelled by a function of the form $+^A_{Q, q}$ or $\times^A_{Q, q}$,
  depending on if $v$ is an addition or multiplication gate. We also add a
  single output gate in $D$ that has as children exactly those gates $(z, q)$
  where $q \in Q_z \cap B$. We define $D$ from $\Phi$ recursively as follows.
  Let $v \in \Phi$.
  \begin{itemize}
  \item If $v$ is a non-constant input gate in $\Phi$ let $(v, 1)$ be an input
    gate in $D$ labelled by the same variable as $v$ and let $(v, 0)$ be a
    $\NOT$-gate with child $(v, 1)$.
  \item If $v$ is a constant gate in $\Phi$ labelled by some field element $q$
    let $(v, q)$ be a constant gate in $D$ labelled by $1$.
  \item Suppose $v$ is an internal gate. Let $Q = \bigcup_{u \in \child{v}}Q_u$.
    For $q \in Q$ let $A_q = \{u \in \child{v} : q \in Q_u\}$. Let $A =
    \biguplus_{q \in Q}A_q$. For each $c \in Q_v$ let $(v, c)$ be a gate in $D$
    such that if $v$ is an addition gate or multiplication gate then $(v, c)$ is
    labelled by $+^A_{Q, c}$ or $\times^A_{Q, c}$, respectively. The labelling
    function $L_{(v, c)} : A \ra \child{v, c}$ is defined for $u \in A$ such
    that if $u \in A_q$ then $L_{(v, c)} (u) = (u, q)$.
  \end{itemize}
  We add one final $\OR$-gate $w$ to form $D$ with $\child{w} = \{(z, q) : q \in
  B \cap Q_z\}$.

  We now show that $D$ is a $G$-symmetric circuit. Let $\sigma \in G$ and $\pi$
  be an automorphism of $\Phi$ extending $\sigma$. Let $\pi' : D \ra D$ be
  defined such that for each gate $(v, c) \in D$, $\pi' (v, c) = (\pi(v), c)$
  and for the output gate $w$, $\pi'(w) = w$. It can be verified by induction
  that $\pi'$ is an automorphism of $C$ extending $\sigma$.
  
  We now show that $\ORB{D} = \ORB{\Phi}$. It suffices to prove that for $v, u
  \in \Phi$ and $c \in Q_v$ that $u \in \orb(v)$ if, and only if, $(u, c) \in
  \orb(v, c)$. The forward direction follows from the above argument
  establishing that $D$ is $G$-symmetric. Let $v, u \in \Phi$ and $c \in Q_v$
  and suppose $(u, c) \in \orb(v, c)$. For each gate $t \in \Phi$ pick some $c_t
  \in Q_t$ such that if $t = u$ or $t = v$ then $c_t = c$ and for all $t_1, t_2
  \in \Phi$, if $Q_{t_1} = Q_{t_2}$ then $c_{t_1} = c_{t_2}$. Let $\pi'$ be an
  automorphism of $D$ such that $\pi'(v, c) = (u, c)$. Let $\pi : \Phi \ra \Phi$
  be defined for $t \in \Phi$ such that $\pi' (t, c_t) = (\pi(t), c_t)$. We now
  show that $\pi$ is an automorphism of $\Phi$, and so $u \in
  \orb(v)$. Since $\pi'$ preserves the labelling on the gates in $D$,
  a simple induction on the depth of the gate in the circuit shows 
  that for all $t \in \Phi$, $Q_{t} = Q_{\pi(t)}$ and so $c_{\pi(t)} = c_t$. Let
  $t, t' \in \Phi$ and suppose $\pi(t) = \pi (t')$. Then $\pi'(t, c_t) =
  (\pi(t), c_t) = (\pi(t), c_{\pi(t)}) = (\pi(t'), c_{\pi(t')}) = (\pi(t'),
  c_{t'}) = \pi'(t', c_{t'})$, and so $(t, c_t) = (t', c_{t'})$ and $t = t'$. It
  follows that $\pi$ is injective, and so bijective. Let $t, s \in \Phi$. Then
  $t \in \child{s} \iff (t, c_t) \in \child{s, c_s} \iff \pi'(t, c_t) \in
  \child{\pi'(s, c_s)} \iff (\pi(t), c_t) \in \child{\pi(s), c_s} \iff \pi(t)
  \in \child{\pi(s)}$. The first and last equivalences follow from the
  construction of the circuit. The remaining conditions for $\pi$ to be an
  automorphism can be easily verified.

  Let $M \in \{0, 1\}^{X}$. We now show by induction that for all $v \in \Phi$
  and $c \in Q_v$, $\Phi[M](v) = c$ if, and only if, $D[M](v, c) = 1$. Let $v
  \in \Phi$. If $v$ is an input gate then the claim holds trivially. Suppose $v$
  is an internal gate and let $c \in Q_v$. Suppose $v$ is an addition gate. Then
  $(v, c)$ is labelled by the function $+^A_{Q, c}$ where $Q = \bigcup_{u \in
    \child{v}} Q_u$, for $q \in Q$, $A_q = \{u \in \child{v} : q \in Q_u\}$, and
  $A = \biguplus_{q \in Q}A_q$. Then
  \begin{align*}
    \Phi[M](v) = c &\iff \sum_{u \in \child{v}} \Phi[M](u) = c 
                     \iff \sum_{q \in Q} \vert \{u \in  \child{v} : \Phi[M](u) = q\}\vert \cdot q = c\\
                   &\iff \sum_{q \in  Q} \vert \{u \in A_q : D[M](u, q) = 1\}\vert \cdot q = c \\
                   &\iff \sum_{q \in Q} \vert \{u \in A_q : L^M_{(v, c)}(u) = 1\} \vert \cdot q = c\\
                   &\iff D[M](v, c) = 1 
  \end{align*}
  A similar argument suffices if $v$ is a multiplication gate. It follows that
  $D[M](w) = 1$ if, and only if, there exists $c \in B$ such that $D[M](z, c) =
  1$ if, and only if, $\Phi[M] \in B$.

  We define $C$ from $D$ by replacing each internal gate $(v, c) \in D$ labelled
  by some $F \in \BA^\ff$ with the rigid strictly $G$-symmetric threshold
  circuit $C(F)$ computing $F$ defined in Lemma~\ref{lem:partition-symmetric}.
  $C$ computes the same function as $D$.  We now argue that $\ORB{C}
  = \ORB{D}$.  Suppose that some gate $g$ in $C(F)$ corresponding to a
  gate $(v,c)$ in $D$ is mapped by an automorphism of $C$ to a gate
  $g'$ in $C(F')$ corresponding to $(v',c')$ in $D$.  Since each
  $C(F)$ has a unique OR gate, it must be the case that the OR gate in
  $C(F)$ then maps to the OR gate in $C(F')$ and so we have an
  isomorphism between $C(F)$ and $C(F')$.  The fact that  $C(F)$ is
  rigid and  \emph{strictly} partition
  symmetric ensures that the isomorphism respects the partition on the
  input and so the circuits compute the same function, i.e. $F = F'$.
  We can conclude that the only
  automorphisms of $C$ are those that are obtained from automorphisms
  of $D$.  Thus, $\ORB{C} = \ORB{D} = \ORB{\Phi}$.
\end{proof}

%% file: sections-journal/supports.tex
Lower bounds that have been established for symmetric Boolean circuits are based on showing lower bounds on \emph{supports} in such circuits.  In this section, we review the connection between the orbit size of circuits, the size of supports and the counting width of graph parameters computed by such circuits.  We improve on the known connection between support size and orbit size to show that it can be used to obtain exponential lower bounds.  We begin by reviewing the definition of supports.

\begin{definition}
  Let $C$ be a rigid $G$-symmetric circuit with variables $\{x_{i, j}: i, j \in
  [n]\}$. We say a set $S \subseteq [n]$ is a \emph{support} of a gate $g$ if
  $\stab_G(S) \leq \stab_G(g)$.
\end{definition}

Let $\consp(g)$ be the minimum size of a support of a gate $g$. Let $\SP(C)$ be
the maximum size of $\consp(g)$ for $g$ a gate in $C$.  We refer to $\SP(C)$ as the \emph{support size} of $C$.

Upper bounds on the orbit size of a square symmetric circuit yield upper bounds on its support size.  Indeed, it was shown in~\cite[Theorem~4]{AndersonD17} that circuit families of size at most $s = \mathcal{O}(2^{n^{1/3}})$ have supports of size at most $\mathcal{O}(\frac{\log s}{\log n})$.   This result was extended to orbit size $s = \mathcal{O}(2^{n^{1-\epsilon}})$ for arbitrary positive $\epsilon$ in \cite[Theorem~1]{AtseriasDO19}.  The result there is stated in terms of the
\emph{size} of the circuit rather than its orbit size.  However, the
proof easily yields the bound for orbit size.  These results
immediately yield that polynomial-size families of symmetric circuits
have $\mathcal{O}(1)$ support size.  It also implies that a linear
lower bound on support size yields a lower bound of
$\Omega(2^{n^{1-\epsilon}})$ on orbit size.  It was this relationship
that was used to obtain lower bounds on the size of symmetirc circuits
for the permanent in the early version of this paper~\cite{DawarW20}.
Here we improve the lower bound by showing that a linear lower bound
on support size implies an exponential lower bound on orbit size, in
Theorem~\ref{thm:support-theorem} below.  First we recall the
following theorem.

\begin{theorem}[{\cite[Theorem 4.10]{DawarW22}}]
  Let $C$ be a rigid square symmetric Boolean circuit with order $n > 8$. For
  every $1 \leq k \leq \frac{n}{4}$ if the maximum size of an orbit in $C$ is
  bounded by ${n \choose k}$ then each gate in $C$ has a support of size less
  than $k$.
  \label{thm:support-theorem-general}
\end{theorem}

Theorem~\ref{thm:support-theorem-general} should be understood as a
restatement of {\cite[Theorem 4.10]{DawarW22}} using the language of
this paper.  In~\cite{{DawarW22}} we dealt with a more general notion
of circuits where individual gates could be labelled by functions that
are not fully symmetric.  What are called circuits with
\emph{injective labels} and \emph{unique extensions} in that paper,
restricted to the circuits we consider here, are exactly the rigid circuits.

We now extract from
Theorem~\ref{thm:support-theorem-general} an asymptotic relationship between
orbits and supports. 

\begin{theorem}
  Let $(C_n)_{n \in \nats}$ be a family of rigid square symmetric Boolean circuits over the threshold basis. If
  $\ORB{C_n} = 2^{o(n)}$ then $\SP(C_n) = o(n)$.
  \label{thm:support-theorem}
\end{theorem}
\begin{proof}
  Let $k$ be the least value such that $\ORB{C_n}  \leq {n \choose k}$.  By the assumption that $\ORB{C_n} = 2^{o(n)}$, we have that $k$ is $o(n)$.  Indeed, otherwise there is a constant $c$ with $0 < c < \frac{1}{2}$, such that $k -1 \geq cn$ for infinitely many $n$.  And since ${n \choose l} \geq (\frac{n}{l})^{l}$ for all $l$ it follows that ${n \choose {k-1} } \geq (\frac{n}{cn})^{cn} > 2^{cn}$.  Since $k$ is the \emph{least} value such that $\ORB{C_n}  \leq {n \choose k}$, it follows that $\ORB{C_n} \geq 2^{cn}$ for infinitely many $n$, contradicting the assumption that $\ORB{C_n} = 2^{o(n)}$.

From $k = o(n)$ it follows that for all large enough $n$, $k \leq \frac{n}{4}$ and so, by Theorem~\ref{thm:support-theorem-general}, $\SP(C_n) \leq k$ and therefore $\SP(C_n) = o(n)$ as required.
\end{proof}

We now use the connection between support size and counting width
established in~\cite{AndersonD17}.  Indeed, Theorem~6
of~\cite{AndersonD17} asserts that a query on relational structures
(e.g.\ a graph property) is decidable by a family of square symmetric
circuits with polynomial orbit size if, and only if, it is definable
in $C^{\omega}_{\infty\omega}$, an infinitary logic with counting
quantifiers.  It is known that definability in this logic is the same
as having bounded counting width.  Moreover, the proof
of~\cite[Theorem 6]{AndersonD17} establishes this by showing that a
circuit of support size $k$ translates into a formula with $O(k)$ variables.  Thus, if a class of
graphs $\mathcal{C}$ is decidable by a family of symmetric circuits $(C_n)_{n
  \in \nats}$ with supports of size at most $k(n)$ then $\mathcal{C}$ has
counting width $\mathcal{O}(k)$. This, along with
Theorem~\ref{thm:support-theorem}, immediately yields the following.

\begin{theorem}
  Let $\mathcal{C}$ be a class of graphs decidable by a family of square
  symmetric Boolean circuits with threshold gates and with orbit size $2^{o(n)}$, then
  $\mathcal{C}$ has counting width $o(n)$.
  \label{thm:cw-graphs}
\end{theorem}

The statement of Theorem~\ref{thm:cw-graphs} does not make mention of the rigidity condition. This suffices, as from~\cite[Lemma~7]{AndersonD17} we have that any symmetric Boolean circuit over the threshold basis may be converted into an equivalent \emph{rigid} symmetric circuit with only a linear increase in size.  It is easily seen from the proof of that lemma that the conversion does not increase orbit size.


For a field $\ff$ and a graph parameter $\mu$ with values in $\ff$, we say that $\mu$ is computed by a family of $\ff$-arithmetic circuits $(C_n)_{n \in \nats}$ if the inputs to $C_n$ are labelled by the variables $x_{ij}$ for $i,j \in [n]$ and,  given the adjacency matrix of a graph $\Gamma$ on its inputs, $C$ computes $\mu(\Gamma)$.

\begin{corollary}
  If a graph parameter $\mu$ is computed by a square symmetric family of arithmetic circuits with orbit size $2^{o(n)}$, then the counting width of $\mu$ is $o(n)$.
  \label{cor:arithmetic-supports-square}
\end{corollary}
\begin{proof}
  Let $k$ be the counting width of $\mu$.  Then, by definition, we can find for each $n \in  \nats$ a pair of graphs $\Gamma_n$ and $\Delta_n$ with at most $n$ vertices such that
  $\Gamma_n \equiv^{k(n) - 1} \Delta_n$ but $\mu (\Gamma_n) \neq \mu(\Delta_n)$.  Let $B_n =
  \{\mu(\Gamma_n)\}$.  Then, by Theorem~\ref{thm:arithmetic-to-boolean}  there is a family of
  square symmetric circuits with threshold gates of orbit size $2^{o(n)}$
  that decides for a graph $\Gamma$
  whether $\mu (\Gamma) \in B_n$.   It follows from Theorem~\ref{thm:cw-graphs} that
  the counting width of this decision problem is $o(n)$.  Since the counting
  width of this decision problem is, by choice of $\Gamma_n$, $k$, it follows that
  $k = o(n)$.
\end{proof}

%% file: sections-journal/permanent-lower-bound.tex
In this section we establish exponential lower bounds on the size of symmetric
arithmetic circuits for the permanent.  We state the result for square
symmetric arithmetic circuits over fields of characteristic zero in
Section~\ref{sec:char-zero} and show how it can be derived from a
lower bound on the counting width of the number of perfect matchings.
The bulk of the section is the construction in
Section~\ref{sec:construction} establishing this counting width lower
bound.  In Section~\ref{sec:positive-char},  we explain how the
argument extends to fields of positive characteristic other than two,
but at the expense of making the stronger requirement that the
circuits are matrix symmetric.  Finally, in
Section~\ref{sec:equivariant} we make a comparison of our lower bounds
with lower bounds on equivariant determinantal representations of the permanent.

\subsection{Characteristic Zero}\label{sec:char-zero}
\begin{theorem}\label{thm:lowerbound-square}
  There is no family of square symmetric arithmetic circuits over any
  field $\ff$ of
  characteristic $0$ of orbit size $2^{o(n)}$ computing $\{\PERM_n\}$.
\end{theorem}

Our proof of this result establishes something stronger. We actually show that
there is no family of symmetric arithmetic circuits of orbit size $2^{o(n)}$
that computes the function $\perm(M)$ for matrices $M \in \ff^{n \times n}$.
Clearly, a circuit that computes the polynomial $\PERM_n$ also computes this
function. Theorem~\ref{thm:lowerbound-square} is proved by showing lower bounds on the
counting widths of functions which determine the number of perfect matchings in
a bipartite graph.

For a graph $\Gamma$ let $\mu(\Gamma)$ be the number of perfect matchings in $\Gamma$.  Our construction establishes a linear lower bound on the counting width of $\mu(\Gamma)$.  Indeed, it also shows a linear lower bound on the counting width of $(\mu(\Gamma) \bmod p)$
for all odd values $p$.
Thus, we aim to prove the following.

\begin{theorem}\label{thm:matching}
  There is, for each $k \in \nats$, a pair of balanced bipartite graphs $X$ and
  $Y$ with $\mathcal{O}(k)$ vertices, such that $X \equiv^k Y$, and $\mu(X) -
  \mu(Y) = 2^l$ for some $l >0$.
\end{theorem}

Before giving the proof of Theorem~\ref{thm:matching} we show how
Theorem~\ref{thm:lowerbound-square} now follows.

\begin{proof}[Proof of Theorem~\ref{thm:lowerbound-square}]
  By Theorem~\ref{thm:matching}, we have, for each $k$, a pair of graphs $X$ and $Y$
  with $\mathcal{O}(k)$ vertices such that  $X
  \equiv^k Y$ and $\mu(X) \neq \mu(Y)$ and hence $\mu(X)^2 \neq \mu(Y)^2$.  Thus, the counting width of $\mu^2$  is $\Omega(n)$.
  Suppose that there is a family of square symmetric arithmetic circuits over a
  field of characteristic $0$ with orbit size $2^{o(n)}$ computing
  $\{\PERM_n\}$.  Then, since the permanent of the adjacency matrix of a bipartite graph $G$ is exactly $\mu(\Gamma)^2$, it follows from
  Corollary~\ref{cor:arithmetic-supports-square} that the counting width of the
  $\mu^2$ is $o(n)$, giving a contradiction.
\end{proof}

It is worth noting why we consider the parameter $\mu^2$ rather than $\mu$ itself in the proof above.  The proof of Theorem~\ref{thm:cw-graphs}, relying on~\cite[Theorem 6]{AndersonD17} relates the counting witdth of a class $\mathcal{C}$ to the size of supports in symmetric circuits deciding $\mathcal{C}$.  Specifically, this is proved for circuits whose input is the adjacency matrix of a graph $\Gamma$ and which are symmetric with respect to permutations of the vertices of the graphs.  This is why we need to take the permanent of the adjacency matrix, rather than the biadjacency matrix of the graph $\Gamma$.  We consider this point in more detail in Section~\ref{sec:positive-char} when we consider lower bounds in fields of positive characteristic.

\subsection{Construction}\label{sec:construction}
The construction to prove Theorem~\ref{thm:matching} is an adaptation of a
standard construction by Cai, F\"{u}rer and Immerman~\cite{CFI92} which gives
non-isomorphic graphs $X$ and $Y$ with $X \equiv^k Y$ for arbitrary $k$ (see
also~\cite{DR07}). We tweak it somewhat to ensure that both graphs have perfect
matchings (indeed, they are both balanced bipartite graphs). The main innovation
is in the analysis of the number of perfect matchings the graphs contain.

\paragraph{Gadgets.}
In what follows, $\Gamma = (V,E)$ is always a 3-regular 2-connected graph. From this,
we first define a graph $X(\Gamma)$. The vertex set of $X(\Gamma)$ contains, for each edge
$e \in E$, two vertices that we denote $e_0$ and $e_1$. For each vertex $v\in V$
with incident edges $f,g$ and $h$, $X(\Gamma)$ contains five vertices. One of these
we call the \emph{balance vertex} and denote $v_b$. The other four are called
\emph{inner vertices} and there is one $v_S$, for each subset $S \subseteq
\{f,g,h\}$ of even size. For each $v \in V$, the neighbours of $v_b$ are exactly
the four vertices of the form $v_S$. Moreover, for each $e \in \{f,g,h\}$,
$X(\Gamma)$ contains the edge $\{e_1,v_S\}$ if $e \in S$ and the edge $\{e_0,v_S\}$
otherwise. There are no other edges in $X(\Gamma)$.
\begin{figure}[h]
  \centering{ 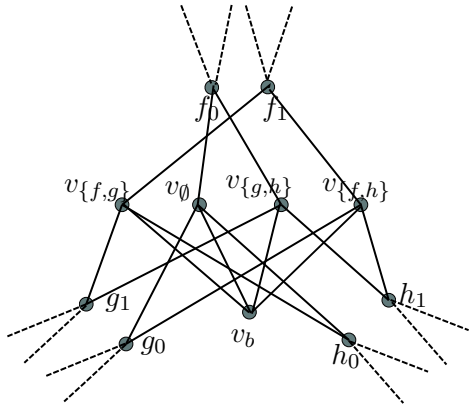
    \caption{A gadget in $X(\Gamma)$ corresponding to vertex $v$ with incident edges
      $f,g,h$}
    \label{fig:gadget}
  }
\end{figure}

The construction of $X(\Gamma)$ from $\Gamma$ essentially replaces each vertex $v$ with
incident edges $f,g$ and $h$ with the gadget depicted in
Figure~\ref{fig:gadget}, where the dashed lines indicate edges whose endpoints
are in other gadgets. The vertices $e_0,e_1$ for each $e \in \{f,g,h\}$ are
shared with neighbouring gadgets.

For any fixed vertex $x \in V$ with incident edges $f,g,h$, the graph $\tX_x(\Gamma)$
is obtained by modifying the construction of $X(\Gamma)$ so that, for the one vertex
$x$, the gadget contains inner vertices $x_S$ for subsets $S \subseteq
\{f,g,h\}$ of odd size. Again, for each $e \in \{f,g,h\}$, $X(\Gamma)$ contains the
edge $\{e_1,v_S\}$ if $e \in S$ and the edge $\{e_0,v_S\}$ otherwise.  Equivalently, we could describe this by saying that in this gadget, we interchange the roles of $g_0$ and $g_1$.

If we remove the balance vertices $v_b$, the graphs $X(\Gamma)$ and $\tX_x(\Gamma)$ are
essentially the Cai-F\"{u}rer-Immerman (CFI) graphs associated with $\Gamma$. The
balance vertex $v_b$ is adjacent to all the inner vertices associated with $v$
and so does not alter the automorphism structure of $X(\Gamma)$ (or $\tX_x(\Gamma)$) at
all. Nor do these vertices alter any other essential properties of the CFI
construction. In particular, since $\Gamma$ is connected, we have the following
lemma. Though it is standard, we include a proof sketch.
\begin{lemma}\label{lem:v-irrelevant}
  For any $x,y \in V$, $\tX_x(\Gamma)$ and $\tX_y(\Gamma)$ are isomorphic.
\end{lemma}
\begin{proof}[Proof (sketch)]
  Note that the gadget corresponding to a vertex $v$ as in
  Figure~\ref{fig:gadget} admits automorphisms that swap $e_0$ and $e_1$ for any
  two edges $e$ incident on $v$. Now, let $x=v_0,v_1,\ldots,v_t = y$ be a simple
  path from $x$ to $y$ in $\Gamma$. We obtain an isomorphism from $\tX_x(\Gamma)$ to
  $\tX_y(\Gamma)$ by interchanging $e_0$ and $e_1$ for all edges on this path, and
  extending this to the induced automorphisms of the gadgets corresponding to
  $v_1,\ldots,v_{t-1}$.
\end{proof}
With this in mind, we refer simply to the graph $\tX(\Gamma)$ to mean a graph
$\tX_x(\Gamma)$ for some fixed $x$, and we refer to $x$ as the special vertex of $\Gamma$.

By known properties of the CFI construction, we also have the following
(see~\cite[Theorem~3]{DR07}).
\begin{lemma}\label{lem:cfi}
  If the treewidth of $\Gamma$ is greater than $k$, then $X(\Gamma) \Cequiv{k} \tX(\Gamma)$.
\end{lemma}

The purpose of the balance vertices is to change the structure of the perfect
matchings. Indeed, if we consider the subgraph of $X(\Gamma)$ that
excludes the balance vertices, it is easily seen that this contains no perfect
matchings. It is a bipartite graph where one part contains the $4|V|$ inner
vertices and the other part contains the $2|E| = 3|V|$ edge vertices and so no
perfect matching is possible. But, $X(\Gamma)$ is a bipartite graph where in one part
we have the $4|V|$ inner vertices and in the other the $3|V|$ edge vertices
along with the $|V|$ balance vertices. In short, this is a $4$-regular bipartite
graph and so contains perfect matchings. We next analyse the structure of the
set of such perfect matchings. In particular, we show that $X(\Gamma)$ and $\tX(\Gamma)$
contain different numbers of perfect matchings.

In the sequel, we write $X$ to denote either one of the graphs $X(\Gamma)$ or
$\tX(\Gamma)$, $V(X)$ to denote its vertices and $E(X)$ to denote its edges. We
continue to use $V$ and $E$ for the vertices and edges of $\Gamma$. Also, for each $v
\in V$, we write $I_v$ to denote the set of four inner vertices in $X$
associated with $v$.

\paragraph{Non-Uniform Matchings.}
Let $M \subseteq E(X)$ be a perfect matching in $X$. For each $v \in V$ and $e
\in E$ incident on $v$, we define the projection $p^M(v,e)$ of $M$ on $(v,e)$ to
be the value in $\{0,1,2\}$ which is the number of edges between $\{e_0,e_1\}$
and $I_v$ that are included in $M$. These satisfy the following equations:
$$p^M(u,e) + p^M(v,e) = 2 \text{ for each edge } e = \{u,v\} \in E; \text{ and }$$
$$p^M(v,f) + p^M(v,g) + p^M(v,h) = 3 \text{ for each vertex } v \in V \text{ with incident edges } f,g,h.$$
The first of these holds because $M$ must include exactly one edge incident on
each of $e_0$ and $e_1$. The second holds because $M$ must include an edge
between $v_b$ and one vertex of $I_v$. Thus, the three remaining vertices in
$I_v$ must be matched with vertices among $f_0,f_1,g_0,g_1,h_0,h_1$.

One solution to the set of equations is obtained by taking the constant
projection $p^M(v,e) = 1$ for all such pairs $(v,e)$. Say that a matching $M$ is
\emph{uniform} if $p^M(v,e) = 1$ everywhere and \emph{non-uniform} otherwise.

\begin{lemma}\label{lem:non-uniform}
  The number of non-uniform matchings in $X(\Gamma)$ is the same as in $\tX(\Gamma)$.
\end{lemma}
\begin{proof}
  It suffices to prove that for any non-constant projection $p$, the number of
  matchings $M$ with $p^{M} = p$ is the same for both $X(\Gamma)$ and $\tX(\Gamma)$. For
  then, taking the sum over all possible projections gives the result. So, let
  $p$ be a non-constant projection. Then, for some edge $e = \{u,v\} \in E$, we
  have $p(u,e) = 2$ and $p(v,e) = 0$. Then, let $X(\Gamma)^-$ and $\tX(\Gamma)^-$ be the
  subgraphs of $X(\Gamma)$ and $\tX(\Gamma)$ respectively obtained by removing the edges
  between $\{e_0,e_1\}$ and $I_v$. It is clear that any matching $M$ in $X(\Gamma)$
  with $p^{M} = p$ is also a perfect matching in $X(G)^-$, and similarly for
  $\tX(\Gamma)$. However, $X(\Gamma)^-$ and $\tX(\Gamma)^-$ are isomorphic. This follows by an
  argument analogous to the proof of Lemma~\ref{lem:v-irrelevant}. Since $\Gamma$ is
  2-connected, there is a path $s = uv_1\cdots v_{t-1}x$ from $u$ to the special vertex $x$ that does
  not involve the edge $e$. We can then define an isomorphism from $X(\Gamma)^-$ to
  $\tX(\Gamma)^-$ by mapping $e_0$ to $e_1$, for each edge $f$ on the path $s$, mapping
  $f_0$ to $f_1$ and extending this using the induced automorphisms of the
  gadgets corresponding to $v_1,\ldots,v_{t-1}$. We conclude that the numbers of
  such matchings are the same for both.
\end{proof}

Now, we aim to show that the number of uniform matchings of $X(G)$ is different
to that of $\tX(\Gamma)$. For this, it is useful to first analyse the orientations of
the underlying graph $\Gamma$.

\paragraph{Orientations.}
An \emph{orientation} of $\Gamma$ is a directed graph obtained from $\Gamma$ by assigning
to each edge $\{u,v\} \in E$ a direction, either $(u,v)$ or $(v,u)$. There are
exactly $2^{|E|}$ distinct orientations of $\Gamma$. We say that a vertex $v \in V$
is \emph{odd} with respect to an orientation $\arr{\Gamma}$ of $\Gamma$ if it has an odd
number of incoming directed edges and \emph{even} otherwise. For an orientation
$\arr{\Gamma}$ of $\Gamma$, we write $\odd{\arr{\Gamma}}$ for the set of its odd vertices. We
say that the orientiation $\arr{\Gamma}$ is \emph{odd} if $|\odd{\arr{\Gamma}}|$ is odd,
and we say it is \emph{even} otherwise.

\begin{lemma}\label{lem:orientation-parity}
  If $|V|/2$ is even, then all orientations of $\Gamma$ are even. If $|V|/2$ is odd,
  then all orientations of $\Gamma$ are odd.
\end{lemma}
\begin{proof}
  Note that since $\Gamma$ is $3$-regular, $3|V| = 2|E|$, so $|V|$ is always even.
  Moreover, $|V|/2$ is even if, and only if, $|E|$ is. For an orientation
  $\arr{\Gamma}$, let $\inc{v}$ denote the number of edges incoming to the vertex
  $v$. Then, $|E| = \sum_v \inc{v}$. But, $\sum_v \inc{v} \equiv |\odd{\arr{\Gamma}}|
  \pmod 2$.
\end{proof}

Thus, we say that a graph $\Gamma$ is \emph{odd} if $|E|$ is odd, and hence all
orientations of $\Gamma$ are odd, and $\Gamma$ is \emph{even} if $|E|$ is even and hence
all orientations of $\Gamma$ are even.

\begin{lemma}\label{lem:transitive-parity}
  If $\Gamma = (V,E)$ is even then for every set $S \subseteq V$ with $|S|$ even,
  there is an orientation $\arr{\Gamma}$ of $\Gamma$ with $\odd{\arr{\Gamma}} = S$. Similarly
  if $\Gamma = (V,E)$ is odd, then for every set $S \subseteq V$ with $|S|$ odd,
  there is an orientation $\arr{\Gamma}$ of $\Gamma$ with $\odd{\arr{\Gamma}} = S$.
\end{lemma}
\begin{proof}
  It suffices to show, for any set $S \subseteq V$ and any pair of vertices $u,v
  \in V$, if there is an orientation $\arr{\Gamma}$ of $\Gamma$ with $\odd{\arr{\Gamma}} = S$,
  then there is also an orientation $\arr{\Gamma}'$ with $\odd{\arr{\Gamma}'} = S
  \symmdiff \{u,v\}$. Now, consider any simple path from $u$ to $v$ in $\Gamma$ and
  let $\arr{\Gamma}'$ be the orientation obtained from $\arr{\Gamma}$ by reversing the
  direction of every edge on this path.
\end{proof}

Indeed, we can say more.

\begin{lemma}\label{lem:linear}
  For every set $S \subseteq V$ with $|S| = |E| \pmod 2$, there are exactly
  $2^{|V|/2 + 1}$ distinct orientations $\arr{\Gamma}$ with $\odd{\arr{\Gamma}} = S$.
\end{lemma}
\begin{proof}
  Let $A$ be the $V \times E$ incidence matrix of the graph $\Gamma$. This defines a
  linear transformation from the vector space $\ff_{2}^{E}$ to $\ff_{2}^{V}$.
  The additive group of $\ff_{2}^{E}$ has a natural action on the orientations
  of $\Gamma$: for a vector $\pi \in \ff_{2}^{E}$, and an orientation $\arr{\Gamma}$,
  define $\pi\arr{\Gamma}$ to be the orientation obtained from $\arr{\Gamma}$ by changing
  the orientation of each edge $e$ with $\pi(e) = 1$. Indeed, fixing one
  particular orientation $\arr{\Gamma}$, the action generates all orientations and
  gives a bijective correspondence between the vectors in $\ff_{2}^{E}$ and the
  orientations of $\Gamma$. Similarly, the additive group of $\ff_{2}^{V}$ has a
  natural action on the powerset of $V$: for a vector $\sigma \in \ff_{2}^{V}$
  and a set $S \subseteq V$, let $\sigma S$ be the set $S \symmdiff \{v \mid
  \sigma(v) = 1\}$. Again, for any fixed set $S$, this action generates all
  subsets of $V$ and gives a bijection between $\ff_{2}^{V}$ and the powerset
  of $V$.

  Then, it can be seen that $\odd{\pi \arr{\Gamma}} = (A \pi) \odd{\arr{\Gamma}}$. Indeed,
  if $v \in V$ is a vertex with incident edges $f,g,h$, then $(A\pi)(v) =
  \pi(f)+\pi(g)+\pi(h) \pmod 2$. In other words $(A\pi)(v) = 1$ just in case the
  direction of an odd number of edges incident on $v$ is flipped by $\pi$. Thus,
  the set of vertices $\{v \mid (A\pi)(v) = 1\}$ are exactly the ones that
  change from being odd to even or vice versa under the action of $\pi$, i.e.
  $\{v \mid (A\pi)(v) = 1\} = \odd{\arr \Gamma} \symmdiff \odd{\pi \arr{\Gamma}}$ for any
  orientation $\arr \Gamma$.

  Fixing a particular orientation $\arr{\Gamma}$, the action of $\ff_{2}^{E}$
  generates all orientation $\pi\arr{\Gamma}$, and $A$ maps this to the collection of
  all sets $\odd{\arr \Gamma} \symmdiff \odd{\pi \arr{\Gamma}}$. Then, by
  Lemmas~\ref{lem:orientation-parity} and~\ref{lem:transitive-parity} the image
  of $A$ consists of exactly the set of vectors with an even number of $1$s.
  Hence, the image of $A$ has dimension $|V|-1$ and so its kernel has size
  $2^{|E|}/2^{|V|-1}$. Since $|E| = 3|V|/2$, this is $2^{|V|/2+1}$. By
  linearity, the pre-image of any vector $v$ in the image of $A$ has exactly
  this size. Thus, for each even size set $T \subseteq V$, there are exactly
  $2^{|V|/2+1}$ vectors $\pi \in \ff_{2}^{E}$ with $\odd{\pi \arr \Gamma} = T
  \symmdiff \odd{\arr \Gamma}$.
\end{proof}

\paragraph{Matchings in Gadgets.}
Any uniform perfect matching $M$ of $X$ induces an orientation of $\Gamma$, which we
denote $\arr{\Gamma}^M$ as follows: any edge $e=\{u,v\} \in E$ is oriented from $u$
to $v$ in $\arr{\Gamma}^M$ if $M$ contains an edge between $e_0$ and a vertex in
$I_u$ and an edge between $e_1$ and a vertex in $I_v$.

Furthermore, every orientation arises from some perfect matching. To see this,
consider again the gadget in Figure~\ref{fig:gadget}. This has eight subgraphs
induced by taking the vertices $\{v_b\} \cup I_v$, together with exactly one
vertex from each of the sets $\{f_0,f_1\}$, $\{g_0,g_1\}$ and $\{h_0,h_1\}$. We
claim that each of these eight subgraphs contains a perfect matching. Indeed, it
suffices to verify this for the two cases $S = I_v \cup \{v_b\} \cup
\{f_0,g_0,h_0\}$ and $T = I_v \cup \{v_b\} \cup \{f_0,g_0,h_1\}$ as the other
six are obtained from these by automorphisms of the gadget. In what follows, we
also write $S$ and $T$ for the subgraphs of the gadget in
Figure~\ref{fig:gadget} induced by these sets.

$S$ has exactly four perfect matchings:
$$
\begin{array}{cccc}
  f_0 - v_{\emptyset} & g_0 - v_{\{f,h\}} & h_0 - v_{\{f,g\}} & v_b - v_{\{g,h\}} \\
  f_0 - v_{\{g,h\}}  & g_0 -  v_{\emptyset}  & h_0 - v_{\{f,g\}}  & v_b - v_{\{f,h\}}  \\
  f_0 - v_{\{g,h\}}  & g_0 - v_{\{f,h\}}  & h_0 - v_{\{f,g\}}  & v_b - v_{\emptyset}\\
  f_0 - v_{\{g,h\}}  & g_0 - v_{\{f,h\}}  & h_0 - v_{\emptyset}  & v_b - v_{\{f,g\}}.
\end{array}
$$

$T$ has exactly two perfect matchings:
$$
\begin{array}{cccc}
  f_0 - v_{\emptyset} & g_0 - v_{\{f,h\}} &  h_1 - v_{\{g,h\}}  & v_b - v_{\{f,g\}}  \\
  f_0 - v_{\{g,h\}}  & g_0 -  v_{\emptyset}   & h_1 - v_{\{f,h\}}  & v_b -  v_{\{f,g\}}. 
\end{array}
$$

Hence, for any orientation $\arr{\Gamma}$, we get a matching $M \subseteq X$ with
$\arr{\Gamma}^M = \arr{\Gamma}$ by choosing one matching from each gadget. To be precise,
for each vertex $v \in V$, define the \emph{relevant subgraph} of $X$ at $v$ to
be the subgraph induced by $I_v \cup \{v_b\}$ along with the vertices $e_1$ for
each edge $e$ incoming at $v$ in $\arr{G}$ and $e_0$ for each edge $e$ outgoing
at $v$ in $\arr{\Gamma}$. In $X(\Gamma)$, the relevant subgraph at $v$ is isomorphic to
$S$ if $v$ is even in $\arr{\Gamma}$ and it is isomorphic to $T$ if $v$ is odd in
$\arr{\Gamma}$. The same is true for all vertices in $\tX(\Gamma)$, apart from the special
vertex $x$. For this one, the relevant subgraph is isomorphic to the induced subgraph on $S$ if $x$ is
odd and to $T$ if $x$ is even. In either case, we get a perfect matching $M$
with $\arr{\Gamma}^M = \arr{\Gamma}$ by independently choosing exactly one matching in
each relevant subgraph. There are $4$ such choices when the relevant subgraph is
like $S$ and $2$ choices when it is like $T$.

\paragraph{Uniform Matchings.}
It follows that for any orientation $\arr{\Gamma}$ of $\Gamma$, the number of uniform
perfect matchings $M$ of $X(\Gamma)$ with $\arr{\Gamma}^M = \arr{\Gamma}$ is
$2^{|\odd{\arr{\Gamma}}|}4^{|V|-|\odd{\arr{\Gamma}}|}$. The number of uniform perfect
matchings in $\tX(\Gamma)$ depends on whether the special vertex $x$ is odd in
$\arr{\Gamma}$ or not. If it is, the number is
$2^{|\odd{\arr{\Gamma}}|-1}4^{|V|-|\odd{\arr{\Gamma}}|+1}$ otherwise it is
$2^{|\odd{\arr{\Gamma}}|+1}4^{|V|-|\odd{\arr{\Gamma}}|-1}$. Thus, if we denote the number
of uniform perfect matchings in $X$ by $\# MX$, then we have
$$
\begin{aligned}
  \# M X(\Gamma) & = \sum_{\arr{\Gamma}} 2^{|\odd{\arr{\Gamma}}|}4^{|V|-|\odd{\arr{\Gamma}}|}
\end{aligned}
$$
where the sum is over all orientations of $\Gamma$. Then, by Lemma~\ref{lem:linear},
$$
\begin{aligned}
  \# M X(\Gamma) & = 2^{|V|/2+1}\sum_{S\subseteq V \; : \; |S| \equiv |E| \pmod 2}
  2^{|S|}4^{|V|-|S|}.
\end{aligned}
$$
By the same token,
$$
\begin{aligned}
  \# M \tX(\Gamma) & = 2^{|V|/2+1}\sum_{S\subseteq V\; : \; |S| \not\equiv |E| \pmod
    2} 2^{|S|}4^{|V|-|S|}.
\end{aligned}
$$

We aim to show that $\# M X(\Gamma)$ and $\# M \tX(\Gamma)$ are different.  Let $P_m$
denote the number $\sum_{S\subseteq [2m] : |S| \text{even}} 2^{|S|}4^{2m-|S|}$
and $Q_m$ denote the number $\sum_{S\subseteq [2m] : |S| \text{odd}}
2^{|S|}4^{2m-|S|}$.
\begin{lemma}\label{lem:pandq}
  For all $m \geq 1$, $P_m - Q_m = 4^m$.
\end{lemma}
\begin{proof}
  We have
  \begin{align*}
    P_m - Q_m & = \sum_{S\subseteq [2m] : |S| \text{even}} 2^{|S|}4^{2m-|S|} - \sum_{S\subseteq [2m] : |S| \text{odd}} 2^{|S|}4^{2m-|S|} \\
              & = \sum_{S \subseteq [2m]} (-1)^{|S|} 2^{|S|}4^{2m-|S|} \\
              & = 4^{2m} \sum_{S \subseteq [2m]}(\frac{-2}{4})^{|S|} \\
    & = 4^{2m}(1-\frac{1}{2})^{2m} = 4^m.
  \end{align*}
  

\end{proof}

\paragraph{Proof of Theorem~\ref{thm:matching}}
By a standard expander graph construction (e.g.~\cite{Ajtai94}), for any $k$, we
can find a $3$-regular graph $\Gamma$ with treewidth at least $k$ and $2n =
\mathcal{O}(k)$ vertices. Then $X(\Gamma)$ and $\tX(\Gamma)$ both have $\mathcal{O}(k)$
vertices and by Lemma~\ref{lem:cfi} we have $X(\Gamma) \Cequiv{k} \tX(\Gamma)$. Moreover,
$X(\Gamma)$ and $\tX(\Gamma)$ have the same number of non-uniform perfect matchings by
Lemma~\ref{lem:non-uniform}. The number of uniform matchings is $2^{n+1}P_n$ in
one case and $2^{n+1}Q_n$ in the other (which is which depends on whether $n$ is
even or odd). Either way, $ |\mu(X(\Gamma)) - \mu(\tX(\Gamma))| = 2^{3n+1}$, which is a
power of $2$ as required.

\subsection{Positive Characteristics}\label{sec:positive-char}

Theorem~\ref{thm:lowerbound-square} gives a lower bound for square-symmetric circuits computing the permanent in characteristic zero, which contrasts neatly with the upper bound for the determinant established in Theorem~\ref{thm:det-upper-bound}.  We now briefly sketch the lower bounds that our method yields for computing the permanent in positive characteristic.  The short statement is that we can get exponential lower bounds for all odd characteristics, but only with respect to a more stringent symmetry requirement---namely \emph{matrix symmetry}.

Theorem~\ref{thm:matching} establishes a lower bound on the counting width of $\mu$---the number of perfect matchings in a graph.  The theorem also establishes a lower bound on $(\mu \bmod p)$, the number of perfect matchings modulo $p$ for any odd value of $p$.  This is because for the graphs $X$ and $Y$ obtained from the theorem, we have $\mu(X) - \mu(Y) = 2^l$ and so $\mu(X) \not\equiv \mu(Y) \pmod p$ for any odd $p$.
\begin{corollary}\label{cor:modp-width}
  For any odd $p$, the counting width of the number of perfect matchings modulo $p$ of a bipartitioned graph with $n$ vertices is $\Omega(n)$.
\end{corollary}
However, we do not have a lower bound on the counting width of $(\mu^2 \bmod p)$.It is quite possible that, for the graphs $X$ and $Y$ of Theorem~\ref{thm:matching} we have $\mu(X)^2 \equiv \mu(Y)^2 \pmod p$.  This is the reason why Theorem~\ref{thm:lowerbound-square} is only formulated for characteristic zero.

The reason that we have to use $\mu^2$ in the proof of
Theorem~\ref{thm:lowerbound-square} has to do with our use of the
connection between the counting width of a class $\mathcal{C}$ of
relational structures and the orbit size of a circuit family deciding
membership in $\mathcal{C}$ as established in~\cite{AndersonD17},
which we use as a black box to get Theorem~\ref{thm:cw-graphs}.  This
connection between counting width and orbit size of circuits is
established in~\cite{AndersonD17} specifically for circuits taking
relational structures as input and which are symmetric under the
action of permutations of the elements.  In the context of graphs, this means it applies to circuits taking the adjacency matrix of a graph $\Gamma = (V,E)$ as input and symmetric under all permutations of $V$.  For such circuits, it establishes that if $\Gamma$ and $\Delta$ are two graphs on vertex set $[n]$ with $\Gamma \equiv^k \Delta$, then their adjacency matrices cannot be distinguished by a $\sym_n$ circuit of small, i.e.\ $2^{o(k)}$ orbit size.  From this, we cannot directly obtain lower bounds for circuits that take the \emph{biadjacency} matrix of a graph as input.  To do this, we have to look inside the black box of Theorem~\ref{thm:cw-graphs}, relating counting width to circuits.

Consider a bipartite graph $\Gamma= (V,E)$ with bipartition $V= A \cup B$,
where $A$ and $B$ both have $n$ elements.  If we identify both sets
$A$ and $B$ with the set $[n]$ (equivalently, if we fix a bijection
between $A$ and $B$), then the biadjacency matrix $\mathcal{B}_\Gamma$ of
$\Gamma$ can be seen as the adjacency matrix of a directed graph $\hat{\Gamma}$
on vertex set $[n]$ with an arc $(i,j)$ whenever there is an edge
between $i \in A$ and $j \in B$.  It then follows directly from the
results of~\cite{AndersonD17} that if we have a pair of bipartite graphs $\Gamma$ and $\Delta$ with $\hat{\Gamma} \equiv^k \hat{\Delta}$, then the biadjacency matrices of $\Gamma$ and $\Delta$ cannot be distinguished by small symmetric circuits.  Unfortunately, for the graphs $X$ and $Y$ of Theorem~\ref{thm:matching}, we are not able to prove that $\hat{X} \equiv^k \hat{Y}$. 

The proof that a pair of structures are equivalent with respect to $\equiv^k$ is often given by as a Duplicator winning strategy in the $k$-pebble \emph{bijection game} (see~\cite{DR07}).  The relation between such winning strategies and lower bounds for symmetric circuits is made explicit in~\cite{Dawar2016}.  This has been greatly expanded to a method for proving  lower bounds for $\Gamma$-symmetric circuits for arbitrary groups $\Gamma$ in~\cite{DWitcs22}.  What this means in our context is that to prove that the biadjacency matrices of the bipartite graphs $X$ and $Y$ are not distinguished by small \emph{square-symmetric} circuits, we need to show a Duplicator winning strategy that respects a fixed bijection between the two parts of the bipartition in $X$ and $Y$.  We are not able to do this.  What we do know is that the Duplicator winning strategy that shows $X \equiv^k Y$ does respect the bipartition itself.  In other words, we can expand the graphs $X$ and $Y$ with colours for the sets $A$ and $B$ (the two parts of the bipartition) and these coloured graphs are still equivalent with respect to  $\equiv^k$.  This is sufficient to establish that their biadjacency matrices are not distnguished by \emph{matrix-symmetric} circuits of size $2^{o(k)}$.  Since for a bipartite graph $G$, the permanent of its biadjacency matrix $\perm_{\ff}(\mathcal{B}_\Gamma)$, over a field $\ff$ of characteristic $p$ is exactly $(\mu(\Gamma) \bmod p)$, this allows us to establish our lower bound.
\begin{theorem}\label{thm:lowerbound-matrix}
  There is no family of matrix-symmetric arithmetic circuits over any field of
  odd characteristic of orbit size $2^{o(n)}$ computing
  $\{\PERM_n\}$.
\end{theorem}

\subsection{Equivariant Determinantal Representations}\label{sec:equivariant}

Lower bounds for computing the permanent in symmetric models of computation have previously been established, notably in the work of Landsberg and Ressayre~\cite{LandsbergR16}.  They establish an exponential lower bound on the \emph{equivariant determinantal complexity} of the permanent, specifically over the complex field $\complex$.  In this section we make a brief comparison of our results with theirs.

The determinantal complexity (DC) of a polynomial $p \in \ff[X]$ is defined to be the least $m$ such that there is an $m\times m$ matrix $M$ with entries that are affine linear forms in $X$ such that $\det(M) = p$.  Such a matrix is called a \emph{determinantal representation} of $p$.  It is known~\cite{Valiant79} that every polynomial in $\VP$ has DC that is at most quasi-polynomial.  It follows that an exponential lower bound on the DC of the permanent would show that it is not in $\VP$, separating $\VP$ from $\VNP$.  Indeed, such a bound would 
show that circuits computing $\PERM_n$ must have size at least $2^{n^{\delta}}$ for some positive $\delta$.  On the other hand, an exponential lower bound on the circuit complexity of the permanent would also yield a similar lower bound for its determinental complexity.  To see this note that using an $O(n^3)$ family of circuits for computing $\{\DET_n\}$ and an $m\times m$ determinantal representation $M$ of the permanent, we get an $O(m^3)$ family of circuits computing $\{\PERM_n\}$.  This is obtained by taking the circuit computing the determinant and attaching to its $m^2$ inputs the circuits (of at most $O(n)$ size) computing the affine linear forms that form the entries of $M$.  Hence a $2^{f(n)}$ lower bound on the circuit complexity fo the permanent gives us a $2^{f(n)/3}$ lower bound on its determinantal complexity.

Landsberg and Ressayre establish exponential lower bounds on any \emph{equivariant} determinantal representation of the permanent, that is one that preserves all the symmetries of the permanent function.  This includes not just the
permutations on entries that we consider, but the entire projective symmetry
group.  Our aim is to see how this relates to our lower bounds on symmetric circuit complexity.  Unfortunately, the relationship is not straightforward in either direction because of the different notions of symmetry used and the symmetry-breaking nature of the translation from circuits to determinantal representations.  To make this explicit, we first introduce some definitions.  These are simplified from (and so less general than) those given by Landsberg and Ressayre but suffice to show that our results are incomparable with theirs. 

Formally, consider a homogeneous polynomial $p \in \complex[X]$.  Let
$\GL_X$ denote the group of invertible linear maps on the vector space
$\complex^X$.  In what follows, we identify $\complex^X$ with the set
of linear forms in the variables $X$, so we can write $Al$ for
$A \in \GL_X$ and $l$ a linear form in $X$.  We extend the notation to
affine linear form by the convention that $Am = c + Al$ when $m = c+
l$ for $c \in \complex$ and $l$ a linear form.

For a map $A \in \GL_X$, we write $p(AX)$ to mean the polynomial obtained from $p$ by replacing each variable $x \in X$ by the linear form $Ax$.  We now define the \emph{symmetry group} of $p$ to be the group $\mathcal{S}_p$ of linear maps $A \in \GL_X$ such that $p(AX) = p$.  In particular, when $X = \{x_{ij} \mid i,j \in [n] \}$ we can think of the elements of $\complex^X$ as $n \times n$ matrices and the symmetry group of $\DET_n$ can be identified with the group $\mathcal{S}_{\DET_n} = (\GL_n \times \GL_n)/\complex \rtimes \ZZ_2$.  Here the action of $(A,B) \in  \GL_n \times \GL_n$ takes $V \in \complex^X$ to $AVB^{-1}/\det(AB^{-1})$ and the action of the non-trivial element in $\ZZ_2$ takes $V$ to $V^T$.

Let  $M$ be an $m\times m$ determinantal representation of a polynomial $p$.  For $A \in \GL_X$, write $M^A$ to be the matrix obtained from $M$ by replacing each entry $m$ by $Am$ (where we see each affine linear form $m$ as a polynomial in $\complex[X])$.   We say that $M$ is an \emph{equivariant} determinantal representation of $p$ if for each $A \in \mathcal{S}_p$ there is a $B \in \mathcal{S}_{\DET_n}$ such that $M^A = BM$.  In other words, all symmetries of $p$ extend to symmetries of $M$.    Landsberg and Ressayre prove that any equivariant determinantal representation of $\PERM_n$ must have size $\Omega(4^m)$.

We could ask if this lower bound yields a lower bound for symmetric circuits just as an exponential lower bound on the \emph{determinantal complexity} of the permanent yields a lower bound for its unrestricted circuit complexity.  This would require a translation, along the lines of Valiant~\cite{Valiant79} from symmetric circuits to equivariant determinantal representations.  There is little reason to believe that we could have such a translation.  For one thing, the symmetry requirement for \emph{square-symmetric} circuits is only that they are invariant under the natural action of $\sym_n$ on $\PERM_n$, and this is a rather small subgroup of $\mathcal{S}_{\PERM_n}$.  Secondly, Valiant's translation of circuits to determinantal representations is not symmetry preserving.  Thus, the representations obtained from this translation applied to square-symmetric circuits are not even guaranteed to be equivariant with respect to the action of $\sym_n$ on $\PERM_n$, let alone that of $\mathcal{S}_{\PERM_n}$.

In the other direction, we could ask if our lower bounds for symmetric circuits for the permanent yield any lower bounds for equivariant determinantal complexity, especially in combination with the polynomial upper bound for transpose-symmetric circuits for the determinant proved in Theorem~\ref{thm:det-upper-bound}.  Indeed, given a circuit $C$ of size $s$ computing $\DET_m$ and an $m\times m$ determinantal representation $M$ of $p$, a polynomial on $n$ variables, we obtain a circuit $C'$ computing $p$ of size $s(m)+ O(nm)$, where the second term represents the size of the subcircuits required to compute the affine expressions making up the entries of $M$.  If an equivariant determinantal expression translates to a symmetric circuit, then a symmetric circuit lower bound can be translated to a lower bound on equivariant determinantal complexity.  Since the symmetry conditions for circuits are less restrictive, this seems plausible, but there is a mismatch.

Consider the case when $p$ is $\PERM_n$, and $C$ is the square-symmetric circuit for $\DET_m$ obtained from Theorem~\ref{thm:det-upper-bound}.  For the circuit $C'$ to be square symmetric, we require that the action of $\sym_n$ on the variables $X =\{x_{ij} \mid i,j \in [n]\}$ extends to automorphisms of the circuit.  Since this action gives a subgroup of $\mathcal{S}_{\PERM_n}$ acting on $\PERM_n$, we know that for each $\pi \in \sym_n$ there is a $B \in \mathcal{S}_{\DET_m}$ such that $M^{\pi} = BM$.  If this map $B$ was itself the action of a permutation in $\sym_m$ on the rows and columns of $M$, the square symmetry of $C$ would guarantee that $C'$ was also square-symmetric.  However, the equivariance of $M$ does not enforce this.  So, to state the lower bound on determinantal complexity that we can get from our results, we define an alternative notion of equivariance.

Say that $M$ is \emph{permutation equivariant} if for each $\pi \in
\sym_n$, there is a permutation matrix $B \in \GL_m$ such that
$M^{\pi} = BMB^{-1}$.  Note that this notion is incomparable with
equivariance of $M$.  We have relaxed the requirement by only asking
that permutations $\pi$ in $\mathcal{S}_{\PERM_n}$ extend to
symmetries of $M$, but we have made it more stringent by asking that
the symmetry they extend to is itself a permutation matrix in
$\mathcal{S}_{\DET_m}$.  Here, we identify the permutation matrix $B$
with the element $((B,B)/1,1)$ in $\mathcal{S}_{\DET_m}$ as this yields
the desired permutation action.

We can now state the following corollary of our results.
\begin{corollary}
Any permutation equivariant determinantal representation of $\PERM_n$ has size $2^{\Omega(n)}$.
\end{corollary}

%% file: gadget.pdf_tex
\begingroup%
  \makeatletter%
  \providecommand\color[2][]{%
    \errmessage{(Inkscape) Color is used for the text in Inkscape, but the package 'color.sty' is not loaded}%
    \renewcommand\color[2][]{}%
  }%
  \providecommand\transparent[1]{%
    \errmessage{(Inkscape) Transparency is used (non-zero) for the text in Inkscape, but the package 'transparent.sty' is not loaded}%
    \renewcommand\transparent[1]{}%
  }%
  \providecommand\rotatebox[2]{#2}%
  \newcommand*\fsize{\dimexpr\f@size pt\relax}%
  \newcommand*\lineheight[1]{\fontsize{\fsize}{#1\fsize}\selectfont}%
  \ifx\svgwidth\undefined%
    \setlength{\unitlength}{174.07873441bp}%
    \ifx\svgscale\undefined%
      \relax%
    \else%
      \setlength{\unitlength}{\unitlength * \real{\svgscale}}%
    \fi%
  \else%
    \setlength{\unitlength}{\svgwidth}%
  \fi%
  \global\let\svgwidth\undefined%
  \global\let\svgscale\undefined%
  \makeatother%
  \begin{picture}(1,0.86427334)%
    \lineheight{1}%
    \setlength\tabcolsep{0pt}%
    \put(0,0){\includegraphics[width=\unitlength,page=1]{gadget.pdf}}%
    \put(0.40434076,0.62346931){\color[rgb]{0,0,0}\makebox(0,0)[lt]{\lineheight{0}\smash{\begin{tabular}[t]{l}$f_0$\end{tabular}}}}%
    \put(0.55236888,0.62346931){\color[rgb]{0,0,0}\makebox(0,0)[lt]{\lineheight{0}\smash{\begin{tabular}[t]{l}$f_1$\end{tabular}}}}%
    \put(0.21346247,0.21249671){\color[rgb]{0,0,0}\makebox(0,0)[lt]{\lineheight{0}\smash{\begin{tabular}[t]{l}$g_1$\end{tabular}}}}%
    \put(0.28942423,0.11900527){\color[rgb]{0,0,0}\makebox(0,0)[lt]{\lineheight{0}\smash{\begin{tabular}[t]{l}$g_0$\end{tabular}}}}%
    \put(0.84842498,0.21833989){\color[rgb]{0,0,0}\makebox(0,0)[lt]{\lineheight{0}\smash{\begin{tabular}[t]{l}$h_1$\end{tabular}}}}%
    \put(0.70234463,0.08589375){\color[rgb]{0,0,0}\makebox(0,0)[lt]{\lineheight{0}\smash{\begin{tabular}[t]{l}$h_0$\end{tabular}}}}%
    \put(0.34171244,0.45353846){\color[rgb]{0,0,0}\makebox(0,0)[lt]{\lineheight{0}\smash{\begin{tabular}[t]{l}$v_{\emptyset}$\end{tabular}}}}%
    \put(0.47311823,0.46825384){\color[rgb]{0,0,0}\makebox(0,0)[lt]{\lineheight{0}\smash{\begin{tabular}[t]{l}$v_{\{g,h\}}$\end{tabular}}}}%
    \put(0,0){\includegraphics[width=\unitlength,page=2]{gadget.pdf}}%
    \put(0.68215613,0.46116013){\color[rgb]{0,0,0}\makebox(0,0)[lt]{\lineheight{0}\smash{\begin{tabular}[t]{l}$v_{\{f,h\}}$\end{tabular}}}}%
    \put(0,0){\includegraphics[width=\unitlength,page=3]{gadget.pdf}}%
    \put(0.12385572,0.45964684){\color[rgb]{0,0,0}\makebox(0,0)[lt]{\lineheight{0}\smash{\begin{tabular}[t]{l}$v_{\{f,g\}}$\end{tabular}}}}%
    \put(0,0){\includegraphics[width=\unitlength,page=4]{gadget.pdf}}%
    \put(0.48466549,0.13401069){\color[rgb]{0,0,0}\makebox(0,0)[lt]{\lineheight{0}\smash{\begin{tabular}[t]{l}$v_b$\end{tabular}}}}%
    \put(0,0){\includegraphics[width=\unitlength,page=5]{gadget.pdf}}%
  \end{picture}%
\endgroup%

%% file: sections-journal/conclusion.tex
We have introduced a novel restriction of arithmetic circuits based on a natural
notion of symmetry. On this basis, we have shown a fundamental difference
between circuits for the determinant and the permanent.  The former is computable using a polynomial-size family of
square symmetric circuits, while the latter requires at least exponential-size
families of square symmetric circuits for fields of characteristic $0$.  The lower bound for the permanent can be extended to fields of odd positive characteristic for matrix-symmetric circuits.

There are several ways in which our results could be tightened. The first would
be to show the existence of polynomial-size circuits for computing the
determinant over arbitrary fields.  Our construction for fields of characteristic
zero is based on Le Verrier's method, which does not easily transfer to other
fields as it relies on division by arbitrarily large integers.  There are general
methods for simulating such division on small fields, but it is not clear if any
of them can be carried out symmetrically.  Indeed, there are many other efficient
ways of computing a determinant and it seems quite plausible that some method that
works on fields of positive characteristic could be implemented symmetrically.
It should be noted, however, that Gaussian elimination is not such a method.
Known results about the expressive power of fixed-point logic with counting
(see, e.g.~\cite{Dawar15}) tell us that there is no polynomial-size family of
symmetric circuits that can carry out Gaussian elimination.  On the other hand,
we do know that the determinant, even over finite fields, can be computed by
exactly such a family of Boolean circuits, as shown by Holm~\cite{Holm10}.  It is
when we restrict to \emph{arithmetic} circuits, and also require symmetry, that
the question is open.

There is a corresponding question for the permanent lower bound. That is, can the
lower bound on square symmetric circuits for the permanent be extended to all
fields of odd positive characteristic. This might be done by adapting our
construction to analyse the counting width of the number of cycle covers of
general graphs. Another approach would be to adapt our construction and choose
$\Gamma$ so that the sum of the numbers of perfect matchings in $X(\Gamma)$ and $\tX(\Gamma)$
is a power of two.  This would suffice to establish that $\mu(X(\Gamma))^2$ and $\mu(\tX(\Gamma))^2$ also differ by a power of two.


We could consider more general symmetries. For example, the determinant has
other symmetries besides simultaneous row and column permutations. The
construction we use already yields a circuit which is symmetric not only with
respect to these but also transposition of rows and columns.  However, we could
consider a richer group that allowed for arbitrary even permutations of the rows
and columns.  In recent work~\cite{DWitcs22} we have been able to show, with this rich
group of symmetries, an exponential lower bound for the determinant.  It would be interesting to identify the exact boundary on the spectrum of symmetries between the tractability and the intractability of the determinant.

Finally, it is reasonable to think that even just considering
square-symmetric circuits, there are polynomials in $\VP$ which do
not admit polynomial-size symmetric arithmetic circuits, by analogy with the
case of Boolean circuits.  Can we give an explicit example of such a polynomial?